\documentclass[preprint,11pt,3p,authoryear]{elsarticle}

\usepackage[draft]{changes} % Use 'draft' to show changes and 'final' to show final version
\definechangesauthor[name={default}, color=purple]{default}

\usepackage{natbib}
\usepackage{svg}
\usepackage{url}
\usepackage{amssymb}
\usepackage{amsmath}
\usepackage{gensymb}
\usepackage{booktabs}
\usepackage{multirow}
\usepackage{threeparttable}
\usepackage{graphicx}
\usepackage{caption}
\usepackage{array}
\usepackage{subcaption}
\usepackage{cleveref}
\usepackage{amsthm}
\usepackage{geometry}
\usepackage{tabularray}
\usepackage{setspace}
\usepackage{lineno}
\usepackage{xcolor}

\begin{document}
\doublespacing
% \linenumbers

\begin{frontmatter}

\title{Differentiable graph neural network simulator for forward and inverse modeling of multi-layered slope system with multiple material properties}

\author[label1]{Yongjin Choi}
\author[label1]{Jorge Macedo}
\author[label1]{Chenying Liu}

\affiliation[label1]{organization={School of Civil and Environmental Engineering, Georgia Institute of Technology},
            city={Atlanta},
            postcode={30332}, 
            state={GA},
            country={USA}}

\begin{abstract}

Graph neural network simulators (GNS) have emerged as a computationally efficient tool for simulating granular flows. Previous efforts have been limited to simplified homogeneous geometries characterized only by the friction angle, which does not reflect the complexity of realistic slopes encountered in engineering practice. This study introduces a differentiable GNS framework designed for multi-layered slope systems comprising both forward and inverse modeling components. The forward component relies on a fine-tuned GNS that incorporates both friction angle and cohesion. Its performance is demonstrated through column collapse and multi-layered slope runout simulations, where the GNS replicates multi-material flow dynamics while achieving significant computational speedup over the Material Point Method (MPM). The inverse modeling component leverages the trained GNS, reverse-mode automatic differentiation, and L-BFGS-B optimization to infer material properties from a target runout geometry. Its performance is demonstrated by back-calculating the material strengths that led to failure-induced runout in a dam system composed of multiple materials. Results are obtained within minutes and show good agreement with the target strength values. The framework introduced in this study provides an efficient approach for forward runout assessments and inverse strength back-calculation in realistic slope systems.

\end{abstract}


\end{frontmatter}

\section{Introduction}

\label{sec:intro}
Granular flow geohazards, such as natural landslides and failures of geotechnical slope systems (e.g., water and tailings dams, heap leach pads, landfills), pose significant risks to infrastructure and human safety, as evidenced by numerous well-documented failures (e.g., the Fundao, Cadia, and Brumadinho dam failures) \citep{morgenstern2016failure_fundao,macedo2024mpm-granular,robertson2019failure_feijao}. Runout assessments are key for understanding the dynamics of granular flows in slope systems and developing effective mitigation strategies. In fact, such assessments are becoming a key component in dam engineering \citep{ICMM2019}. Numerical approaches capable of capturing large deformations and the complex mechanics of granular flows have been increasingly used in runout modeling. Examples of recent efforts include the Oso landslide \citep{yerro2019runout_mpm_oso}, the Cadia dam failure \citep{macedo2024runout_mpm_cadia}, and the Caloveto landslide \citep{TRONCONE2022runout_mpm_caloveto}. However, numerical frameworks face notable computational challenges, particularly in large-scale simulations. For example, particle-based methods like the Discrete Element Method (DEM) \citep{staron2005column_collapse_dem, kermani2015column_collapse_dem, cleary2002dem3d} simulate micro-structural particle-to-particle interactions but become computationally prohibitive when modeling large-scale flow scenarios that involve a large number of particles and contact interactions. Conventional continuum-based approaches, such as the Finite Element Method (FEM), struggle with large deformations in granular flows due to mesh distortion \citep{sordo2024hybrid-fem-mpm}. The Material Point Method (MPM) \citep{mast2014mpm-granular, macedo2024mpm-granular, nguyen2020mpm-granular} addresses this limitation by adopting a hybrid Eulerian-Lagrangian approach, but its reliance on repeated mapping between material points and a background grid introduces significant computational overhead. This overhead becomes particularly pronounced in inverse modeling scenarios, which require multiple forward simulations to infer unknown system parameters \citep{choi2024inverse}.

Recent advancements in learned physics simulators~\citep{battaglia2016interaction, battaglia2018inductive, sanchez2020learning} offer promising alternatives to numerical approaches. Among them, Graph Neural Network Simulators (GNS) \citep{sanchez2020learning} have emerged as efficient surrogate models for computationally intensive approaches such as DEM \citep{jiang2024gns_inverse, zhao2025physical}, MPM \citep{choi2024graph, haeri2024gns-mpm, kumar2023gns-inverse-forward-mpm, zhao2025physical}, and Smoothed Particle Hydrodynamics (SPH) \citep{li2022graph}. The GNS represents particles as nodes and their interactions as edges in a graph, enabling the model to learn the underlying physics of granular flow dynamics. This structure supports forward predictions with generalization beyond the training domain, similar to numerical frameworks, while achieving computational speedups of several orders of magnitude compared to advanced numerical techniques \citep{zhao2025physical, choi2024graph}. The GNS has also shown potential in inverse modeling due to its computational efficiency and inherent differentiability \citep{zhao2022learning_pde_constrained, jiang2024gns_inverse, allen2022physical}. Its computational efficiency enables rapid exploration of parameter spaces via repeated forward simulations, which would otherwise be prohibitively expensive. Additionally, its differentiability allows for gradient-based optimization in high-dimensional inverse problems \citep{allen2022physical, jiang2024gns_inverse, choi2024inverse, kumar2023accel}. Despite the discussed GNS potential, prior GNS-based efforts relevant to slope systems and runout modeling have focused on simplified geometries, often limited to homogeneous rectangles/cubes (e.g., \cite{choi2024graph,zhao2025physical}), characterized solely based on a friction angle. These simplifications hinder the applicability of the GNS to slope systems encountered in engineering practice, which typically involve multi-layered geometries and materials characterized by both friction angle and cohesion.

In this study, we present a differentiable GNS framework for forward and inverse modeling of realistic, multi-layered slope systems characterized by friction angle ($\phi$) and cohesion ($c$). We fine-tune a GNS initially trained on frictional materials to incorporate both $\phi$ and $c$, enabling it to simulate frictional and cohesive runout behaviors. We evaluate the framework’s forward modeling performance through column collapse and multi-layered slope runout simulations, benchmarking its predictions against MPM results. For inverse modeling, we integrate the trained GNS with reverse-mode automatic differentiation and the L-BFGS-B (Limited-memory Broyden–Fletcher–Goldfarb–Shanno with Box constraints) optimization algorithm to infer material properties from a target runout geometry. This constrained optimization formulation enables the incorporation of prior knowledge, allowing for efficient strength back-calculation from runout observations. The proposed framework provides a computationally efficient, physics-consistent approach for both forward runout prediction and inverse strength estimation in realistic slope systems.

\section{GNS Framework for Forward and Inverse Modeling
%Methods
}\label{sec:method}

\subsection{
GNS forward modeling
}

We build upon the multi-GPU GNS implementation developed by \cite{choi2024graph} as the foundation for our forward simulation framework. For comprehensive implementation details and architectural specifications, the reader is referred to that work. Here, we briefly summarize the core components of the GNS framework and highlight the modifications introduced in this study to support multi-material modeling.

\begin{figure}[!htbp]
    \centering
    \includegraphics[width=0.7\textwidth]{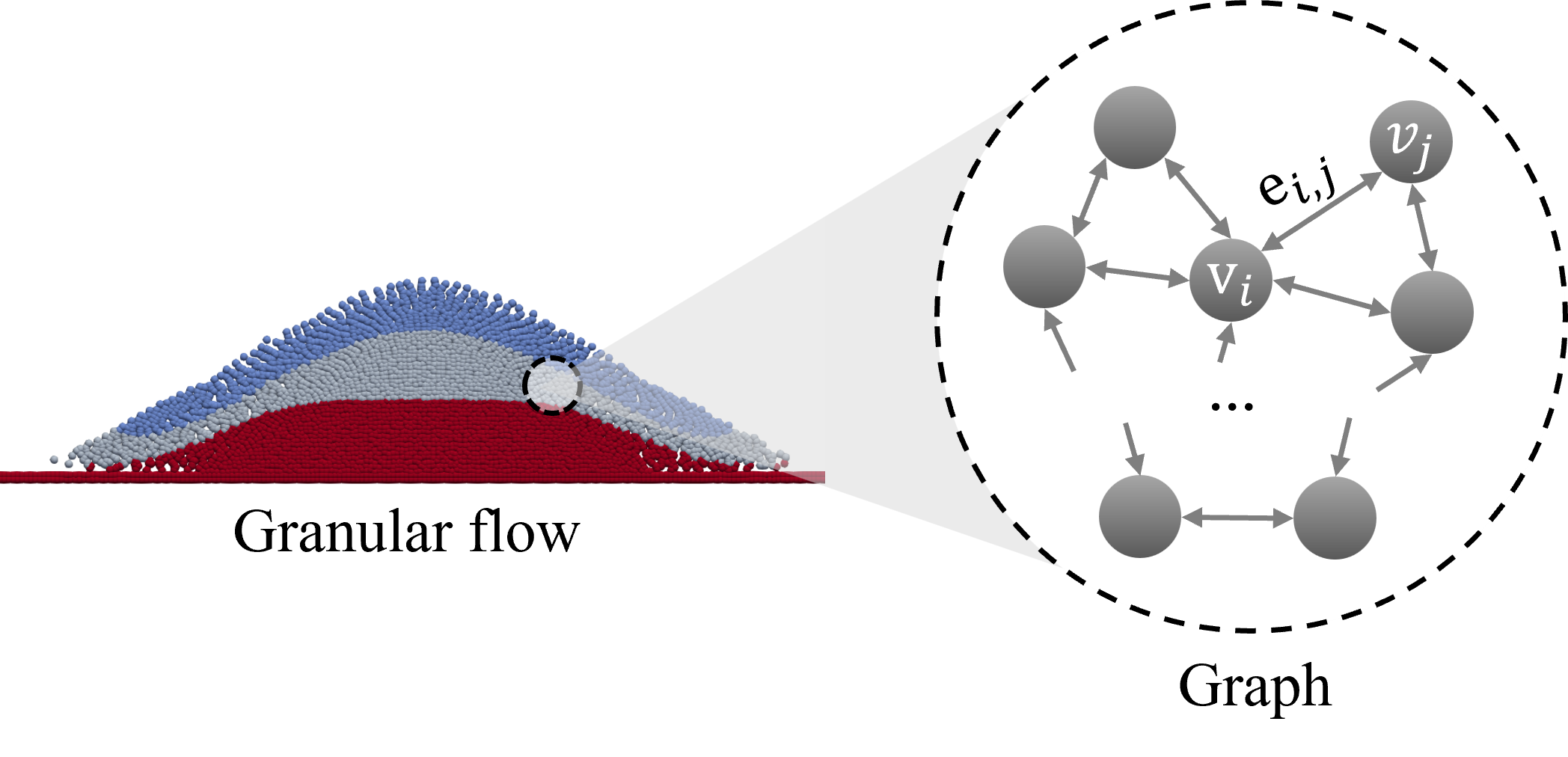}
    \caption{The GNS representation of a granular flow. The graph $G = (V, E)$ consists of a set of vertices $V =  \{ \boldsymbol{v}_1, \boldsymbol{v}_2, \ldots, \boldsymbol{v}_i \}$, and  edges $E = \{ \boldsymbol{e}_{1,1}, \boldsymbol{e}_{1,2}, \ldots, \boldsymbol{e}_{i,j} \} $ connecting the vertices.} 
    \label{fig:graphs}
\end{figure}

The GNS models the physical state of a granular flow as a graph, $ G = (V, E) $ (see \Cref{fig:graphs}), where $ V $ is the set of vertices and $ E $ is the set of edges. Each vertex $ \boldsymbol{v}_i \in V = \{ \boldsymbol{v}_1, \boldsymbol{v}_2, \ldots, \boldsymbol{v}_i \} $ represents a soil particle or a discretized material region (i.e., a material point), while each edge $ \boldsymbol{e}_{i,j} \in E = \{ \boldsymbol{e}_{1,1}, \boldsymbol{e}_{1,2}, \ldots, \boldsymbol{e}_{i,j} \} $ encodes the physical interaction between connected vertices. In our implementation, training data are generated using the Material Point Method (MPM), where vertices correspond to material points and edges capture their physical interactions.

%\subsection{Graph neural network simulator}

\begin{figure}[!htbp]
    \centering
    \includegraphics[width=0.95\textwidth]{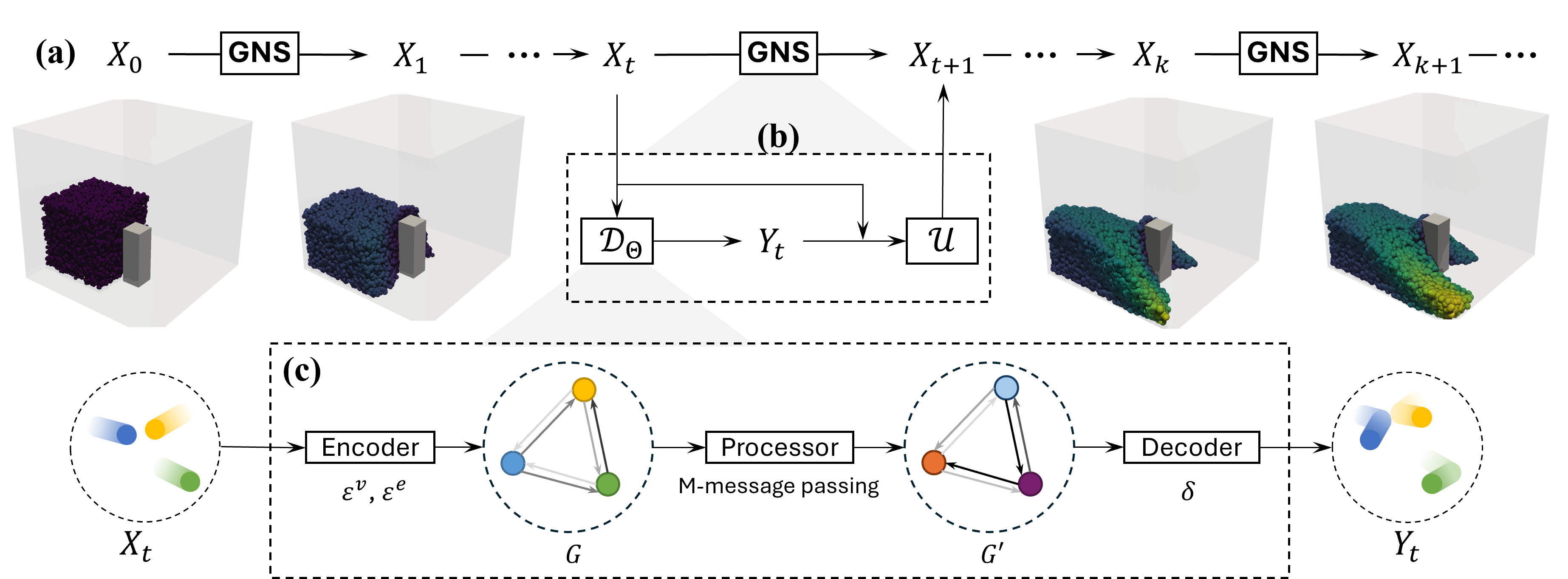}
    \caption{Simulating granular flows using the Graph Neural Network Simulator (GNS). The GNS consists of a learned dynamics model $\mathcal{D}_{\Theta}$ and an update function $\mathcal{U}$ (subfigure b). At each timestep, $\mathcal{D}_{\Theta}$ receives the current physical state $X_t$ and predicts the particle-wise dynamics $Y_t$ based on the encoder-processor-decoder operations (subfigure c). The update function $\mathcal{U}$  then uses $Y_t$ to produce the next state $X_{t+1}$. By iteratively applying this process, the GNS performs autoregressive rollouts ($X_0 \rightarrow \textbf{GNS} \rightarrow X_{1} \rightarrow \textbf{GNS} \rightarrow ... \rightarrow X_{t} \rightarrow \textbf{GNS} \rightarrow X_{t+1}\rightarrow ...$) to simulate the spatio-temporal evolution of granular flow (subfigure a).}
    \label{fig:gns_structure}
\end{figure}

The GNS (\Cref{fig:gns_structure}) predicts the physical state of the system at the next timestep $X_{t+1} = \{\textbf{x}_1^{t+1}, \textbf{x}_2^{t+1}, \ldots, \textbf{x}_i^{t+1}\}$, given the current state $X_t = \{\textbf{x}_1^t, \textbf{x}_2^t, \ldots, \textbf{x}_i^t\}$ as defined in \Cref{eq:gns}. 

\begin{equation}\label{eq:gns}
    X_{t+1} = GNS(X_t)
\end{equation}  

Each particle’s feature vector $\textbf{x}_i^t$ is defined by \Cref{eq:x_feature_vector}:

\begin{equation}\label{eq:x_feature_vector}
    \textbf{x}_i^t = [\textbf{p}_i^{t-k:t}, \textbf{b}_i^t, \textbf{f}, \textbf{w}]
\end{equation}  

Here, $t$ represents timestep, $i$ denotes the particle index, $\textbf{p}_i^{t-k:t}$ contains particle positions from $t-k$ to $t$ (we set $k$=6, following  \cite{choi2024graph}), $\textbf{b}_i^t$ encodes boundary information, $\textbf{f}$ denotes the particle type (e.g., stationary or kinematic), and $\textbf{w}$ encodes material properties. 

The implementation by \citet{choi2024graph} supports only a single material property, i.e., friction angle. In contrast, our study extends the GNS to accommodate multi-material features by modifying the input dimensionality of the relevant components (specifically, the encoder discussed later) based on the number of elements in $\boldsymbol{w}$ (\Cref{eq:x_feature_vector}). This allows the GNS to learn interactions between different material types and provides the flexibility to incorporate additional material parameters as needed. For Mohr-Coulomb materials, $\textbf{w}=[\phi, c]$. 

The GNS follows an autoregressive rollout scheme (\Cref{fig:gns_structure}a), recursively using its own predictions as input to simulate the $k$th timestep of interest (\Cref{eq:gns_rollout}). The simulation timestep $k$ in \Cref{eq:gns_rollout} is not a fixed hyperparameter of the model but is determined by the targeted simulation runout. For the runout analyses, the GNS continues autoregressive predictions until the system reaches equilibrium, capturing the flow process from initiation to the final runout. The simulations presented in the results section typically use 400 to 600 steps to develop a significant runout and a stable final configuration. The GNS operates with a fixed timestep that is directly tied to the physical timestep used in the MPM simulation data sampling for training data. This study samples the MPM simulation at every dt = 0.0375 seconds.

\begin{equation}\label{eq:gns_rollout}
    X_0 \rightarrow \textbf{GNS} \rightarrow X_{1} \rightarrow \textbf{GNS} \rightarrow ... \rightarrow X_{k-1} \rightarrow \textbf{GNS} \rightarrow X_{k}
\end{equation}  

The GNS comprises two main components: a dynamics approximator $\mathcal{D}_{\Theta}$ (Equation~\ref{eq:dynamics_approximator}) and an update function $\mathcal{U}$ (Equation~\ref{eq:update_fn}) (\Cref{fig:gns_structure}b) to perform the computations in \Cref{eq:gns}.

\begin{equation}\label{eq:dynamics_approximator}
    Y_{t} = \mathcal{D}_{\Theta}(X_t)
\end{equation}  

\begin{equation}\label{eq:update_fn}
    X_{t+1} = \mathcal{U}(X_t, Y_t)
\end{equation}  

The dynamics approximator $\mathcal{D}_{\Theta}$, parameterized by learnable weights $\Theta$, follows an encoder, processor, and decoder architecture (\Cref{fig:gns_structure}c). The encoder maps particle states, $X_t$, to a latent graph $G=(V, E)$. The encoder consists of the vertex encoder $\varepsilon^v$ (\Cref{eq:vertex_encoder}) and the edge encoder $\varepsilon^e$ (\Cref{eq:edge_encoder}). 

\begin{equation}\label{eq:vertex_encoder}
    \textbf{v}_i = \varepsilon^v(\textbf{x}_i^t)
\end{equation}  

\begin{equation}\label{eq:edge_encoder}
    \textbf{e}_{i,j} = \varepsilon^e(\textbf{r}_{i,j}^t)
\end{equation}

where, $\textbf{r}_{i,j}^t = \left[\left(\textbf{p}_i^t - \textbf{p}_j^t\right), \ \lVert \textbf{p}_i^t - \textbf{p}_j^t \rVert\right]$, captures relative displacement and distance of material points at the current timestep.

The processor applies $M$ rounds of message passing to propagate information at the graph vertices along the edges of $M$-hop neighbors, returning an updated graph $G'=(V', E')$. This step models energy or momentum transfer between material points. Each step $m = 1,2,...,M$ consists of the following operations.

\begin{enumerate}
    \item \textbf{Message construction:} each edge $(i, j)$ construct a message $\textbf{e}_{i,j}^{(m)}$ using: 
    $$\textbf{e}_{i,j}^{(m)} = \phi^e\left(\textbf{v}_i^{(m-1)}, \textbf{v}_j^{(m-1)}, \textbf{e}_{i,j}^{(m-1)}\right)$$
    where $\phi^e$ is a learnable message function.
    
    \item \textbf{Message aggregation:} each vertex $i$ collects messages from its neighboring vertices:
    $$\bar{\textbf{e}}_i^{(m)} = \sum_{j \in \mathcal{N}(i)} \textbf{e}_{i,j}^{(m)}$$ 
    where $ \mathcal{N}(i) $ denotes the set of neighboring vertices connected to vertex $ i $.
    
    \item \textbf{Vertex update:} each vertex $i$ incorporates the aggregated messages and updates its latent feature:
    $$\textbf{v}_i^{(m)} = \phi^v\left(\textbf{v}_i^{(m-1)}, \bar{\textbf{e}}_i^{(m)}\right)$$ where $\phi^v$ is a learnable vertex update function.
\end{enumerate}

After the message passing, the processor outputs the updated graph $G'=(V', E')$ with $V' = \{\textbf{v}_i^{(M)}\}$ and $E' = \{\textbf{e}_{i,j}^{(M)}\}$. Then, the decoder ($Y_{t} = \text{Decoder}(G')$) extracts the dynamics $Y_t = \{ \textbf{y}_1^t, \textbf{y}_2^t, \ldots, \textbf{y}_i^t \}$ for each particle  using the decoder function $\delta$ (\Cref{eq:vertex_decoder}): 

\begin{equation}\label{eq:vertex_decoder}
    \textbf{y}_i^t = \delta\left(\textbf{v}_i^{(M)}\right)
\end{equation}

The update function $\mathcal{U}$ (\Cref{fig:gns_structure}b) uses $Y_t$ to advance the system state to $X_{t+1}$. $\mathcal{U}$ operates similarly to explicit Euler integration in numerical solvers.

We implement all GNS components--the encoders $\varepsilon^v$, $\varepsilon^e$, processor functions $\phi^e$, $\phi^v$, and decoder $\delta$--using multilayer perceptrons (MLPs) with two hidden layers of 128 units each. Together, these form the learnable parameter set $\Theta$ in $\mathcal{D}_{\Theta}$. We train $\Theta$ by minimizing the mean squared error between the predicted accelerations $Y_t$ and ground truth accelerations $A_t = \{\textbf{a}_1^t, \textbf{a}_2^t, \ldots, \textbf{a}_i^t \}$, obtained from training data generated via MPM simulations using the CB-Geo codebase \citep{kumar2019scalable}.

% Explain the error accumulation

During training, the GNS learns to predict the next state based on the ground truth input sampled from the data. In contrast, during inference, the GNS relies on its own prior predictions, each containing a small error, as inputs for future steps (\Cref{eq:gns_rollout}). These errors accumulate over time, potentially pushing the model inputs outside the distribution seen during training, which could result in inaccuracies.

We mitigate this error accumulation by adopting the random-walk noise injection strategy introduced by \cite{sanchez2020learning}. \cite{sanchez2020learning} reported that noise injection could reduce the rollout error by an order of magnitude compared to the rollout without any noise injection; hence, the strategy is adopted in this study. Interested readers are referred to \cite{sanchez2020learning, pfaff2020meshnet} for further implementation details and discussion of noise sensitivity. Specifically, during training, we perturb the input positions in \Cref{eq:x_feature_vector} with Gaussian noise that accumulates over time, simulating the compounding errors observed during rollout. The corresponding acceleration targets $A_t$ are adjusted so that, after the updating (\Cref{eq:update_fn}), the model still correctly predicts the next position $X_{t+1}$. This strategy enables the model to learn robust correction mechanisms against accumulated error. We use a noise standard deviation of 0.067 m/s for velocity perturbations, selected based on the overall velocity standard deviation in the training dataset. This standard deviation is used to impose noise on the domain material point positions cumulatively. 

\subsection{GNS inverse modeling}

The differentiable GNS framework used for inverse analysis builds on the implementation by \citet{choi2024inverse}. In this study, we extend their framework by incorporating a second-order optimization method, the Limited-memory Broyden-Fletcher-Goldfarb-Shanno algorithm with box constraints (L-BFGS-B) \citep{byrd1995lbfgs-b}, which enables imposing explicit bounds on model parameters during optimization. Below, we describe the differentiable GNS framework and provide a brief overview of the L-BFGS-B optimizer.

\begin{figure}[!htbp]
    \centering
    \includegraphics[width=0.95\textwidth]{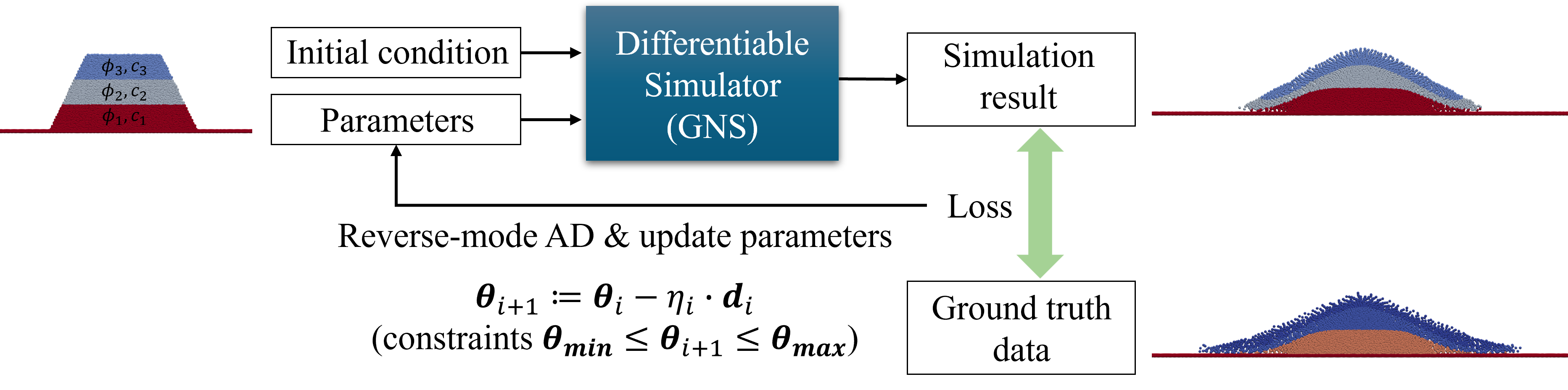}
    \caption{Procedures for solving inverse problems in granular flows with differentiable graph neural network simulator.}
    \label{fig:differentiable_gns}
\end{figure}

\Cref{fig:differentiable_gns} illustrates the differentiable GNS framework for back-calculating the material parameters of selected slope units based on observed runout geometries. The process begins by defining initial conditions and a material parameter set, $\boldsymbol{\theta}$, which are input into the GNS for forward simulations. We then compare the simulation results with ground truth data using a loss function $ \mathcal{L}_{\boldsymbol{\theta}}$, defined as:

\begin{equation}\label{eq:inverse_loss}
    \mathcal{L}_{\boldsymbol{\theta}} = \frac{1}{|\mathcal{T}|} \sum_{t \in \mathcal{T}} \left( P_{\boldsymbol{\theta}}^t - \hat{P}_{\boldsymbol{\theta}}^t \right)^2
\end{equation}

where $\boldsymbol{\theta}$ denotes the set of material parameters, $P_{\boldsymbol{\theta}}^t$ the observed particle positions at time $t$, and $\hat{P}_{\boldsymbol{\theta}}^t$ is the corresponding simulated positions. The term $|\mathcal{T}|$ represents the number of timesteps included in the loss function. Typically, we evaluate the loss at the final timestep (i.e., $|\mathcal{T}| = 1$)--when the slope system reaches equilibrium--as this is the most commonly available observation for engineering applications. However, we may include multiple timesteps ($|\mathcal{T}| > 1$) if available.

We update material parameters using gradient-based optimization, guided by the gradient of the loss function with respect to $\boldsymbol{\theta}$. The parameter update at iteration $i$ follows:

\begin{equation}\label{eq:gd-general}
    \boldsymbol{\theta}_{i+1} = \boldsymbol{\theta}_i - \eta_i \cdot \boldsymbol{d}_i
\end{equation}

Here, $\eta_i$ is the learning rate, and $\boldsymbol{d}_i$ is the update direction. First-order methods such as Stochastic Gradient Descent (SGD) or Adaptive Moment Estimation (ADAM) \citep{kingma2014adam} compute $\boldsymbol{d}_i$ based on the gradient $\nabla f(\boldsymbol{\theta}_i)$. Second-order methods, such as Broyden-Fletcher-Goldfarb-Shanno (BFGS), determine $\boldsymbol{d}_i$ by incorporating curvature information through an approximate Hessian $H = \nabla^2 f(\boldsymbol{\theta}_i)$, estimated from the history of gradients, which improves convergence.

The differentiable GNS framework uses reverse-mode automatic differentiation (AD) \citep{baydin2018ad}, which enables efficient and accurate gradient computation. Reverse-mode AD operates in two phases: a forward pass rolls out the simulation and stores intermediate states (i.e., $X_0, \ X_1, \ \ldots, \ X_k$), and a backward pass propagates gradients through the computational graph using the chain rule, from the final output back to the input. Unlike finite differences, AD computes gradients for all input parameters in a single forward pass, making it highly efficient for high-dimensional inverse problems. 

However, reverse-mode AD imposes high memory demands, particularly in GNS applications involving long simulation rollouts, like slope runout simulations, since intermediate states must be stored at each timestep for backpropagation. To mitigate this memory consumption, we implement gradient checkpointing \citep{chen2016gradcheck}, a technique that reduces memory usage by storing only selected intermediate states, referred to as checkpoints, and recomputing the remaining states on demand during the backward pass. In our implementation, we checkpoint the system state at each simulation timestep. This strategy effectively reduces peak memory consumption at the cost of increased computation, resulting in a 2-3× increase in forward simulation time. Despite this overhead, the differentiable GNS remains significantly faster than high-fidelity solvers \citep{choi2024inverse}, making it suitable for inverse analysis of runout in slope systems. Full implementation details can be found in \cite{choi2024inverse}.

\subsubsection{Constrained optimization with L-BFGS-B}
Back analyses of slope runout failures typically focus on estimating the strength parameters of critical layers involved in the failure mechanism while keeping the properties of non-critical layers fixed (e.g., Lower San Fernando and Las Palmas dams)\citep{olson2001liquefaction, moss2019flow}. Constraining the optimization to bound material properties of noncritical layers with known prior information narrows the search space, enhances solution stability, and prevents non-physical estimates. We employ L-BFGS-B, a quasi-Newton method that achieves fast convergence with low memory usage and supports box constraints on individual parameters. These constraints take the form $\boldsymbol{\theta}_{min} \leq \boldsymbol{\theta} \leq \boldsymbol{\theta}_{max}$, where $\boldsymbol{\theta}_{min}$ and $\boldsymbol{\theta}_{max}$ are the lower and upper bounds for the parameter set $\boldsymbol{\theta}$. The BFGS algorithm approximates the inverse Hessian using historical gradients and parameter updates, avoiding the $O(n^3)$ cost of computing the true Hessian \citep{nocedal2006numerical}. L-BFGS \citep{liu1989lbfgs} extends BFGS by limiting memory usage, retaining only a small window of past updates to reconstruct matrix-vector products on demand. L-BFGS-B \citep{byrd1995lbfgs-b} further enhances this framework by enabling box constraints. During each iteration, L-BFGS-B classifies parameters as either fixed (at constraint bounds) or free (within bounds) using gradient-based projection. It then updates only the free parameters using the L-BFGS update rule. This structure ensures compliance with bounds while maintaining the convergence speed of second-order optimization. By incorporating bounds for strength properties, L-BFGS-B guides the inverse modeling process toward physically constrained solutions, making it well-suited for geotechnical applications involving heterogeneous slope systems. For further algorithmic details of L-BFGS-B, we refer readers to \cite{byrd1995lbfgs-b}, \cite{nocedal2006numerical}.

\section{Training}

We use transfer learning \citep{feng2019transfer, chamangard2022transfer} to reduce the computational cost and data requirements of training the GNS, which would otherwise demand extensive granular flow datasets and specialized high-memory GPUs \citep{choi2024graph}. Transfer learning enables a model to reuse previously learned representations rather than learning entirely from scratch. This approach is particularly effective when expanding the capabilities of a simulator to take into account additional material parameters or previously unmodeled physical processes.

We implement a two-stage training strategy. In the first stage, we train a base GNS model on a dataset in which material behavior depends solely on the friction angle. This phase establishes a pre-trained model that captures the core dynamics of frictional granular flows. In the second stage, we fine-tune this model using a smaller, targeted dataset that incorporates both friction and cohesion. We use this approach since the friction angle is a more fundamental and generic property governing the response of particulate materials, such as soils composing slope systems, which are the focus of this study. Moreover, previous studies \citep{choi2024graph,jiang2024gns_inverse} have established the ability of the GNS to model the dynamics of purely frictional materials. The adopted sequential training considers that the trained GNS can integrate cohesion effects additively while retaining its learned understanding of frictional behavior, which is evaluated in later sections. The following section describes the base dataset used in the initial training and details the subsequent fine-tuning procedure.

\subsection{Datasets}
We use the MPM with a standard MPM scheme \citep{sulsky1995application} and an explicit time integration to generate the training datasets. \Cref{table:training_datasets} summarizes the MPM analysis configurations for the dataset generation and material properties for the base model and fine-tuning stages. While both datasets share consistent MPM settings (element size, number of material points per cell, and timestep), they differ in material properties: the base dataset considers only friction angle, while the fine-tuning dataset includes both friction angle and cohesion. Geostatic stress conditions are applied in all simulations to capture initial slope stress states, and MPM simulations are run. 

\begin{figure}[!htbp]
    \centering
    \includegraphics[width=0.7\textwidth]{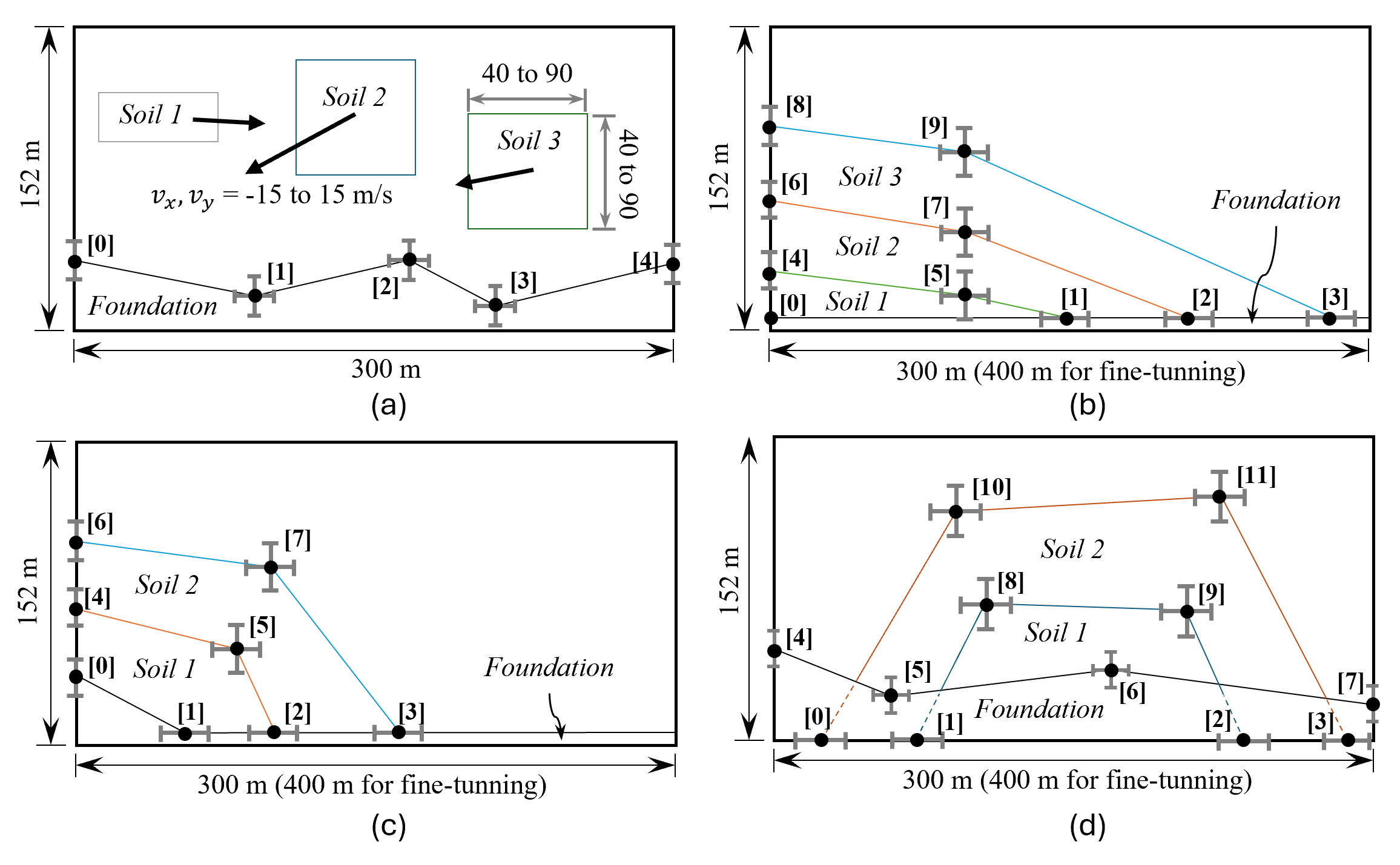}
    \caption{Configuration for randomly generating MPM simulation setups for training data.} The figures are not in scale. For the base dataset, all the configurations from (a) to (d) are included. For the fine-tuning dataset, we exclude (a), and additional column collapse scenarios are included. The numbers in square brackets (e.g., [1], [2], etc.) indicate the indices of the polygon vertices (black dots) that define the geometry. Solid lines connecting these vertices delineate material boundaries. For each simulation, the x- and y-coordinates of the polygon vertices are randomly changed; gray range bars at each vertex illustrate the horizontal and vertical variation. The extent of this positional variation is provided in \Cref{table:appendix_training_data} in \ref{sec:appendix_training_data}. In subfigure (a), the cube length randomly varies from 40 to 90 m, with the black arrows representing the random initial velocity $v_x, v_y$ applied to the cube body, ranging from -15 m/s to 15 m/s. \Cref{fig:appendix_training_data} provides a few example simulation results to facilitate visual understanding.
    \label{fig:train_data_config}
\end{figure}

The base dataset comprises 2,000 MPM simulations generated using randomly defined geometric configurations shown in \Cref{fig:train_data_config}a-d, with each configuration contributing 500 simulations. Polygonal geometries are defined by black dots representing the polygon vertices, each labeled with an index in square brackets (e.g., [1], [2], etc.). The solid lines connecting these vertices delineate material boundaries. Material properties are randomly assigned to each domain based on the value ranges listed in \Cref{table:training_datasets}. For each simulation, the x- and y-coordinates of the polygon vertices are randomly changed; gray range bars at each vertex illustrate the horizontal and vertical variation. The extent of this positional variation is provided in \Cref{table:appendix_training_data} in \ref{sec:appendix_training_data}. In \Cref{fig:train_data_config}a, which simulates the collision of multiple cubes (up to three), black arrows indicate the initial random velocities ($v_x$, $v_y$), which range from -15 to 15 m/s. Cube side lengths vary randomly between 40 and 90 m. In all cases---vertex positions, initial velocities, cube side lengths, and material properties---random sampling is performed using uniform distributions.

The configurations shown in \Cref{fig:train_data_config}b-d are designed to represent multi-layered slope systems and to train the GNS to capture diverse runout modes and material interactions. Specifically, \Cref{fig:train_data_config}b simulates steep natural slopes composed of multiple layers, while \Cref{fig:train_data_config}c represents more gently sloping profiles. \Cref{fig:train_data_config}d models multi-layered systems with both left and right slopes, resembling dams and embankments.

\Cref{fig:train_data_config}a considers a wide range of granular flow behaviors from collisional to frictional regimes, to establish generalizable representations of granular physics. This configuration targets slope systems with significant runout and velocity development, where collisional dynamics are likely. Previous studies (e.g., \cite{sanchez2020learning,choi2024graph,zhao2025physical}) also adopted this configuration to train the GNS for general granular flow simulations. Together, these randomized and layered scenarios expose the GNS to both general and targeted runout behaviors, aiming to enhance its ability to generalize across a range of slope systems. Example simulation results illustrating the deformation behavior for each configuration are presented in \Cref{fig:appendix_training_data}.

The fine-tuning dataset focuses on the scenarios in \Cref{fig:train_data_config}b-d and omits the configuration in \Cref{fig:train_data_config}a to focus the training on multi-layer slopes. Of the 1,000 MPM simulations in the fine-tuning dataset, 250 follow the geometry in \Cref{fig:train_data_config}b, 250 in \Cref{fig:train_data_config}c, and 500 in \Cref{fig:train_data_config}d. We extend the simulation domain width from 300 m (used in the base dataset) to 400 m to better capture potential longer runout trajectories. The fine-tuning stage also considers additional scenarios of column collapse runout (27 simulations) to complement the training. The column widths and heights range between 50 and 100 m.

\begin{table}[]
\centering
\begin{threeparttable}
\caption{Datasets for training base model and model fine-tuning.}
\label{table:training_datasets}
\begin{tabular}{@{}cccc@{}}
\toprule
\multicolumn{2}{c}{\multirow{2}{*}{Property}} & \multicolumn{2}{c}{Dataset} \\ \cmidrule(l){3-4} 
\multicolumn{2}{c}{} & Base & Fine-tuning \\ \midrule
\multirow{6}{*}{MPM configuration} & Boundary (m) & 300 x 152 & 400 x 152 \\
 & Element (m) & \multicolumn{2}{c}{4 x 4} \\
 & \# material points per cell & \multicolumn{2}{c}{4} \\
 & Total \# material points & \multicolumn{2}{c}{6K} \\
 & Data sampling dt & \multicolumn{2}{c}{0.0375} \\ \midrule
\multirow{5}{*}{Material} & $\phi$ ($\degree$) \tnote{a} & 10 to 40 & 3 to 35 \\
 & c (kPa) \tnote{a} & None & 10 to 60 \\
 & E (MPa) & \multicolumn{2}{c}{40} \\
 & $\rho$ ($\text{kg/m}^3$) & \multicolumn{2}{c}{1800} \\
 & $\nu$ & \multicolumn{2}{c}{0.3} \\ \midrule
\multicolumn{2}{c}{\# simulations} & 2000 & 1000 \\ \bottomrule
\end{tabular}%
\begin{tablenotes}
\item[a] Friction is normalized using $\tan(\phi)$ and cohesion divided by 100 kPa to facilitate convergence during training.
\end{tablenotes}
\end{threeparttable}
\end{table}

The generation of the training datasets constitutes a substantial computational effort. Runtimes vary; however, each simulation takes approximately 50 minutes to complete on a 56-core Intel Cascade Lake processor. With 3,000 simulations, the total computational demand amounts to roughly 150,000 minutes. To manage this workload, we leverage the distributed computing capabilities of the Frontera high-performance computing (HPC) cluster at the Texas Advanced Computing Center (TACC), enabling parallel simulation.

\subsection{GNS training}

\begin{figure}[!htbp]
    \centering
    \includegraphics[width=0.7\textwidth]{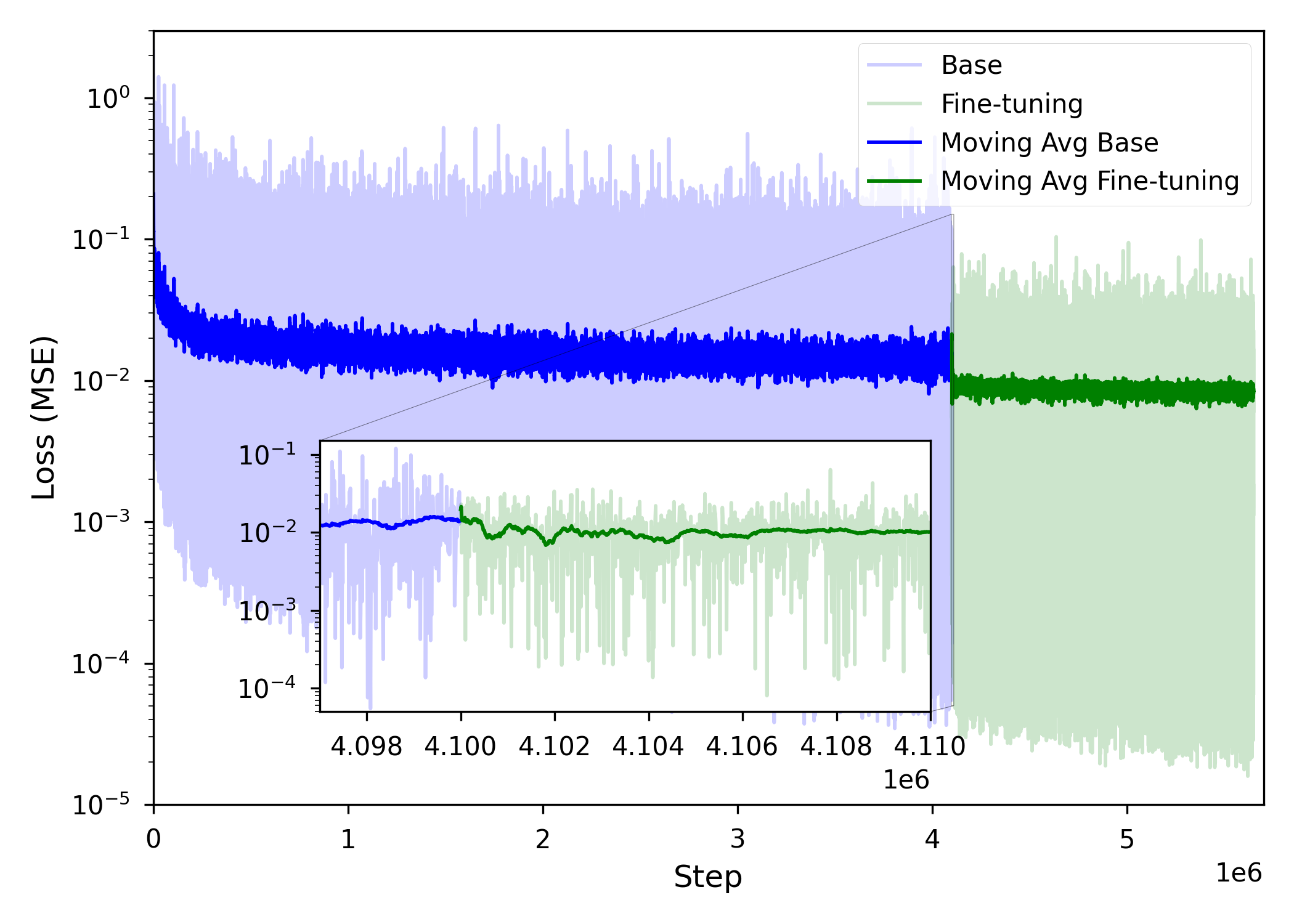}
    \caption{Training loss history. The light blue and dark blue lines represent the raw loss history and its moving average during training on the base dataset. The light green and dark green lines correspond to those during training on the fine-tuning dataset. The inset plot shows the enlarged part of the transition between the base training phase and the fine-tuning phase.}
    \label{fig:train_loss_history}
\end{figure}

We train the GNS on the base dataset for 4.1 million steps prior to fine-tuning. This number reflects the maximum runtime allowable on the HPC system used. While training loss had largely stabilized before completing 4.1 million steps (\Cref{fig:train_loss_history}), we chose to utilize the full computational allocation to maximize model refinements. This approach aligns with previous studies (e.g., \cite{sanchez2020learning,kumar2022minority}), which have suggested that continued training can yield incremental improvements even when the loss decreases only marginally. We use multi-GPU training with two NVIDIA A100 GPU with 40 GB memory on TACC with a batch size of 2 for each GPU. At a training speed of $\sim$7 steps per second, completing 4.1 million steps takes approximately 6-7 days.

\Cref{fig:train_loss_history} shows the training loss history, computed as the mean squared error (MSE) of one-step acceleration predictions from \Cref{eq:dynamics_approximator}. The figure illustrates two distinct phases: base training (light and dark blue for raw and moving-average losses), where only friction is considered, and fine-tuning (light and dark green), where both friction and cohesion are considered.

We initialize training with a learning rate of $lr_{init} =1 \times 10^{-4}$ multiplied by the number of GPUs ($n_{gpu}$) with a batch size of two. During base training, the learning rate follows an exponential decay schedule $lr(i) = lr_{\text{init}} \cdot 0.1^{\,i / 5 \times 10^{6}} \cdot n_{\text{gpu}}$, where $i$ is the training step. At the onset of fine-tuning, training continued from the base training model weights using the reduced learning rate at that training step.

During the base training phase, the loss steadily decreases as the model learns to predict the dynamics of frictional granular flows. At the onset of fine-tuning (see the inset plot showing the enlarged part of the transition between the base training phase and the fine-tuning phase), the loss shows a short surge due to the introduction of the previously unseen material property (cohesion) but quickly drops within $\sim$ 1,000 steps and stabilizes after an additional $\sim$ 7,000 steps. The stabilized loss ($\sim$ 0.009, after around 4.108e6 steps) is lower than the final loss at the end of base training ($\sim$ 0.013, at around 4.1e6 steps). This is largely because the fine-tuning dataset excludes high-velocity collision simulations from \Cref{fig:train_data_config}a, which tend to produce higher MSE values ($\sim$ 0.02) and increase the base training loss. Overall, the loss history indicates an adequate transition from base training to fine-tuning, suggesting that the model successfully incorporates cohesive behavior without notable degradation of its prior knowledge about frictional dynamics.

\section{Performance Assessment and Applications}

\subsection{Forward simulation}

\subsubsection{Granular column collapse with Mohr-Coulomb properties}

We evaluate the performance of the fine-tuned GNS model on Mohr-Coulomb-type materials using the granular column collapse experiment--a well-established benchmark for studying granular flow dynamics under controlled conditions \citep{nguyen2020granular, lajeunesse2005granular}. The setup consists of a vertical column of granular material placed on a flat, rigid surface that collapses under gravity.

The runout dynamics are primarily governed by the column’s initial aspect ratio $a = H_0/L_0$, where $H_0$ and $L_0$ are the initial height and base length, respectively. At low aspect ratios, only the column's edge fails and flows, while the interior remains largely static due to internal friction. At high aspect ratios, most of the column collapses, and the upper portion undergoes near free-fall, resulting in longer runout distances. We quantify runout using the normalized runout distance: $L^n = (L_f - L_0)/L_0$, where $L_f$ is the final runout distance. We assess the GNS performance across three aspect ratios: short ($a =0.5$), intermediate ($a =1.0$), and tall ($a =2.0$), corresponding to initial column sizes of 200 × 100 m, 150 × 150 m, and 100 × 200 m, respectively. Each case is simulated using both GNS and MPM. We consider a range of material parameters: friction angles $\phi = (0, \ 11.25, \ 22.5, \ 33.75, \ 45 ) \degree$ and cohesion values $c = (10, \ 25, \ 40, \ 55, \ 70) \ \text{kPa}$. Of note, we considered test cases with features that were not included during training.
For example, we included different material properties including very low and high friction angles ($\phi=0, 45 \degree$), high cohesion value ($c=70$ kPa), longer simulation durations (600 vs. 400 timesteps), and larger granular bodies (e.g., 200 m column height).% and single-material setups.

\begin{figure}[!htbp]
    \centering
    \includegraphics[width=1.0\textwidth]{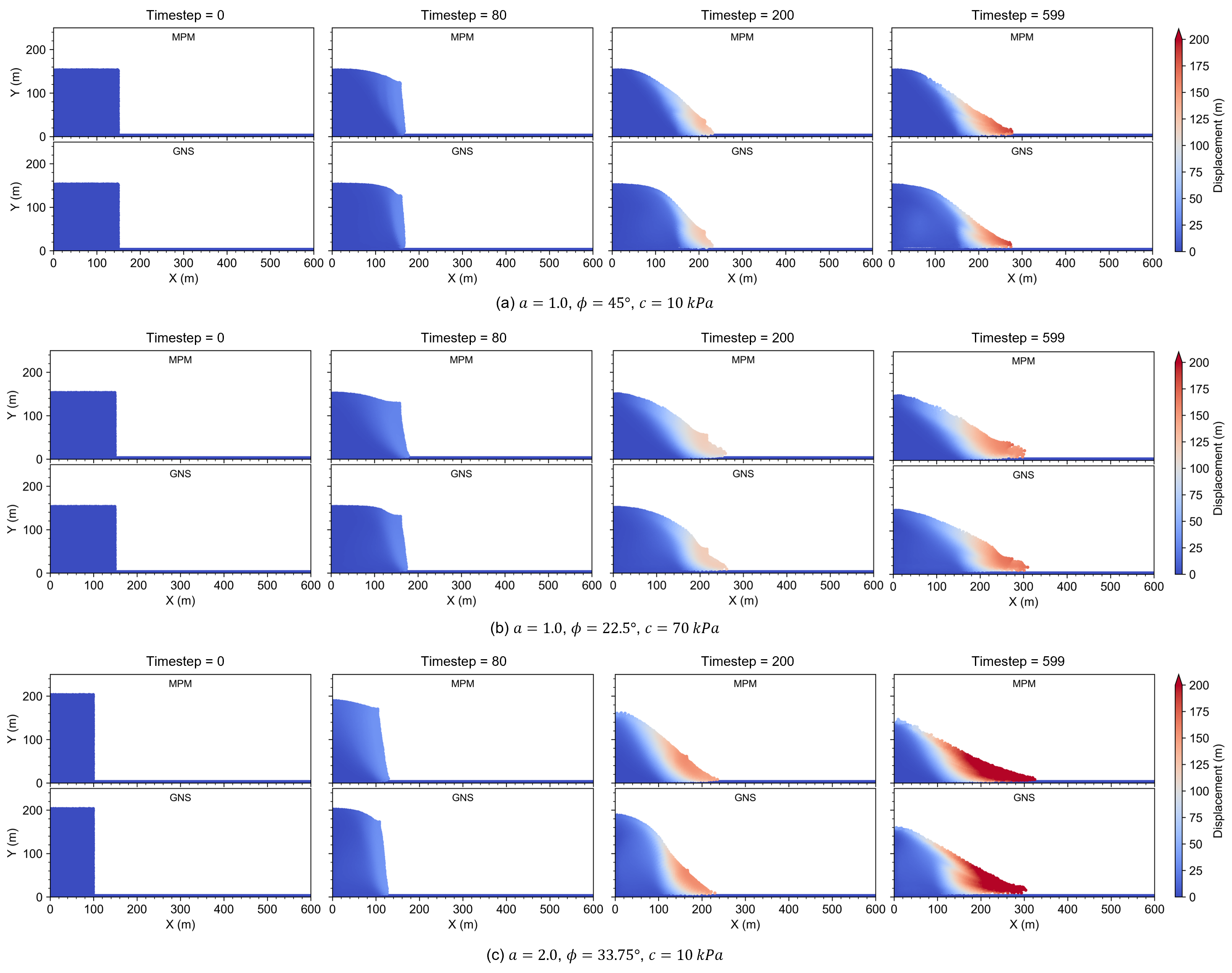}
    \caption{GNS versus MPM granular flow evolution over time. (a) Case with $a = 1.0$, $\phi = 45\degree$, $c = 10~\text{kPa}$, showing a runout error of 1.73\%, ranked at the 25th percentile among all test cases. (b) Case with $a = 1.0$, $\phi = 22.5\degree$, $c = 70~\text{kPa}$, showing a runout error of 7.64\%, ranked at the 50th percentile. (c) Case with $a = 2.0$, $\phi = 33.75\degree$, $c = 10~\text{kPa}$, showing a runout error of 9.27\%, ranked at the 75th percentile.}
    \label{fig:column_collapses}
\end{figure}

\Cref{fig:column_collapses} shows time evolution examples of granular flows in both GNS and MPM simulations. We compute the error as the absolute percentage difference between the normalized runout predicted by the MPM and GNS: $\frac{\left| L^n_{MPM} - L^n_{GNS} \right|}{L^n_{MPM}} \times 100$. \Cref{fig:column_collapses}a presents the case with $a=1.0$, $\phi=45$, $c=10 \ \text{kPa}$) where the GNS predicts a normalized runout of 0.863 compared to 0.848 from MPM, yielding a 1.73\% error, ranked at the 25th percentile among all column collapse test cases. \Cref{fig:column_collapses}b shows the case with $a = 1.0$, $\phi = 22.5\degree$, $c = 70~\text{kPa}$, where the normalized runout from the GNS and MPM is 1.039 and 1.088, respectively, showing a runout error of 7.64\%, ranked at the 50th percentile. \Cref{fig:column_collapses}c shows the case with $a = 2.0$, $\phi = 33.75\degree$, $c = 10~\text{kPa}$, where the normalized runout from the GNS and MPM is 2.316 and 2.102, respectively, showing a runout error of 9.27\%, ranked at the 75th percentile.
In general, the GNS captures the overall runout patterns, runout extent, and geometry evolution adequately. Even considering the 75th percentile error case (\Cref{fig:column_collapses}c ) where the GNS estimates a larger runout, the error is less than 9.27\%, and the overall runout patterns are also captured.

\begin{figure}[!htbp]
    \centering
    \includegraphics[width=1.0\textwidth]{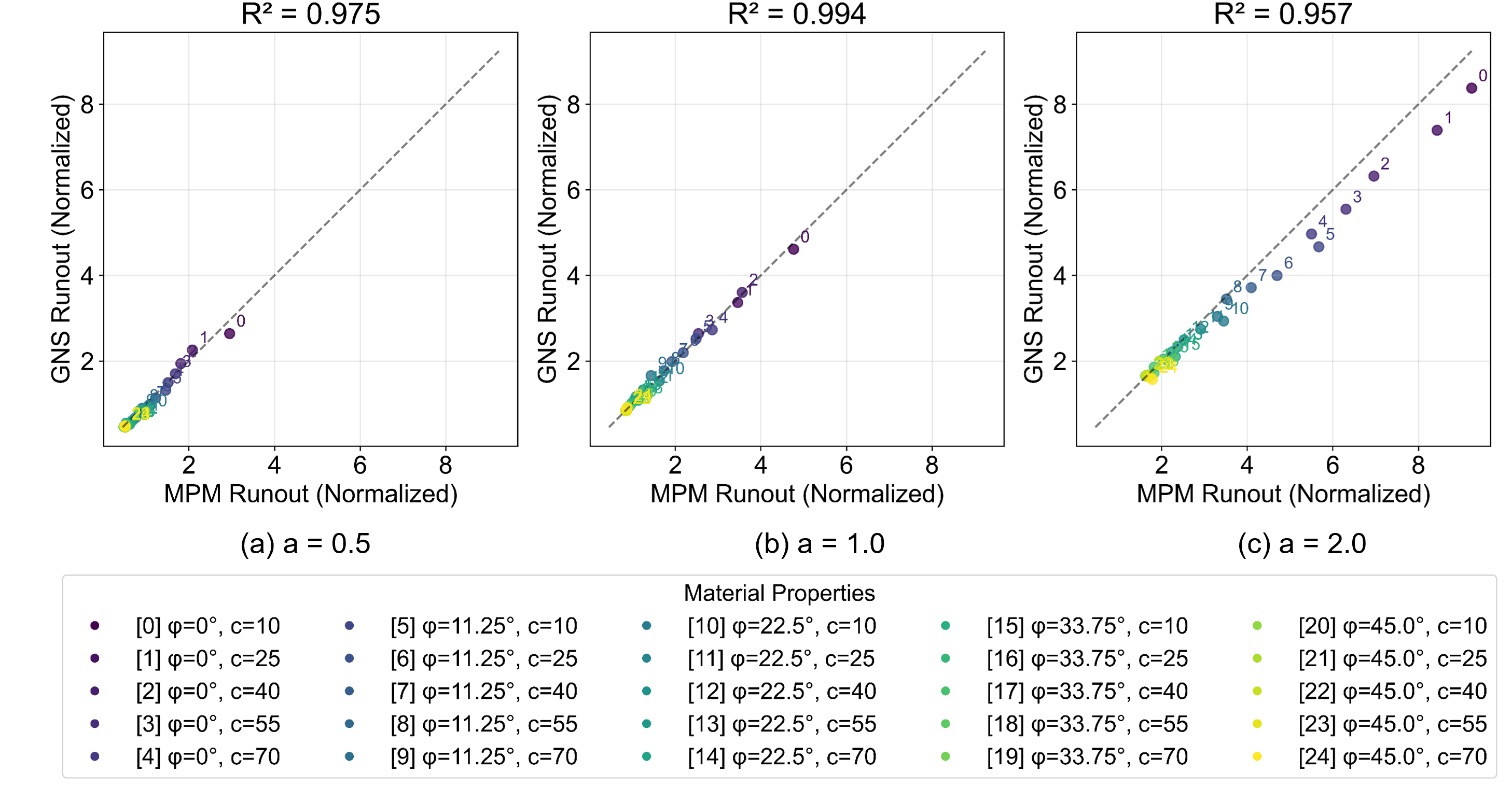}
    \caption{Comparison between normalized runout distances for the granular column collapse test cases between the MPM and GNS with different friction angles ($\phi$) and cohesions ($c$).}
    \label{fig:column_runout_comparison}
\end{figure}

\Cref{fig:column_runout_comparison} presents the comparison between normalized runout distances between the MPM and GNS across all considered combinations of cohesion and friction angle. The three subplots correspond to column configurations with different initial aspect ratios: a = 0.5 (\Cref{fig:column_runout_comparison}a), a = 1.0 (\Cref{fig:column_runout_comparison}b), and a = 2.0 (\Cref{fig:column_runout_comparison}c). Each point represents a unique combination of friction angle and cohesion, with color indicating the material configuration as shown in the legend.

The GNS runout predictions show good agreement with MPM results, closely aligning near the 1:1 line in each plot, yielding coefficients of determination ($R^2$) of 0.975, 0.994, and 0.957 for a = 0.5, 1.0, and 2.0, respectively. \Cref{fig:column_runout_comparison}c (a = 2.0) shows the largest deviations compared to the other cases, particularly for the material property labels 0 to 6, which have zero or near-zero friction angles or low cohesion values. These configurations exhibit a near free-fall, fluid-like behavior characterized by high accelerations and rapid velocities, amplifying the sensitivity of runout predictions. Moreover, this highly dynamic flow regime deviates significantly from the scenarios used for training. These factors likely contribute to the larger runout errors compared to the other shorter columns.

\begin{figure}[!htbp]
    \centering
    \includegraphics[width=1.0\textwidth]{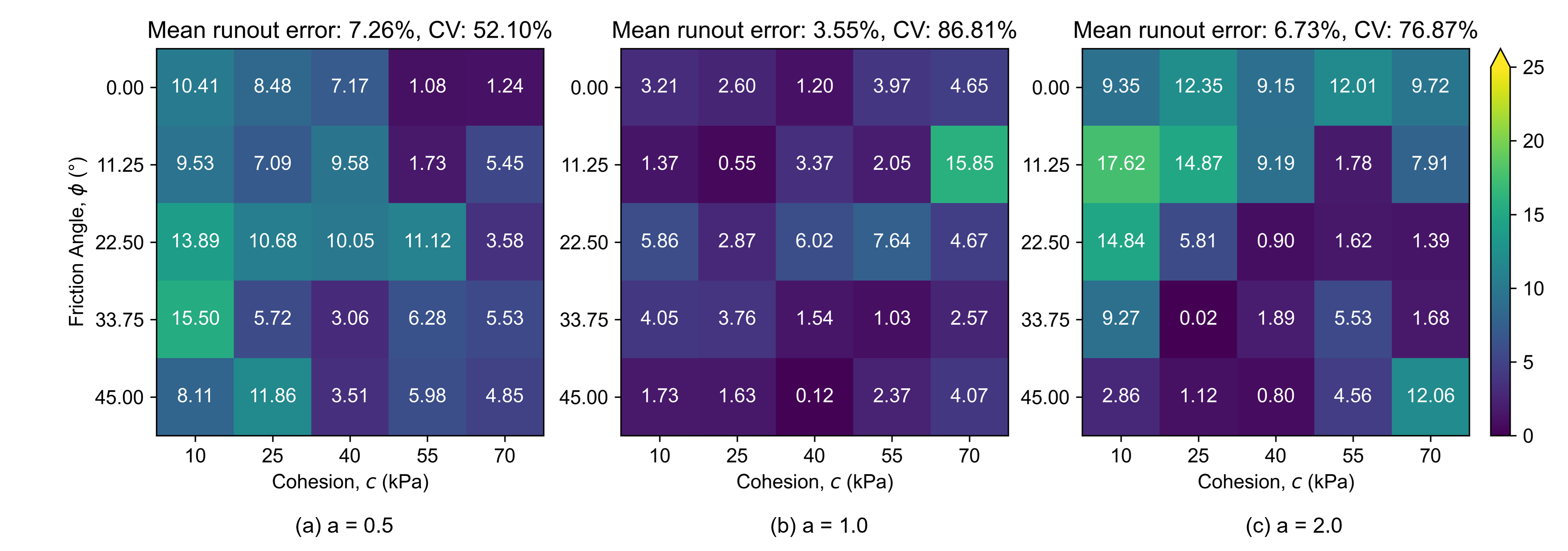}
    \caption{Heatmaps of absolute runout error (\%) between the GNS and MPM across different combinations of friction angle ($\phi$) and cohesion ($c$) for three column aspect ratios: a = 0.5 (left), a = 1.0 (middle), and a = 2.0 (right). The mean and coefficient of variation (CV) are shown in the subfigure title.}
    \label{fig:column_error_heatmap}
\end{figure}

\Cref{fig:column_error_heatmap} presents the absolute percentage runout errors across all test cases shown in \Cref{fig:column_runout_comparison}, categorized by column aspect ratio. The mean runout errors for $a = 0.5$, $a = 1.0$, and $a = 2.0$ are 7.26\%, 3.55\%, and 6.73\%, respectively. Among the three, the $a = 1.0$ case (\Cref{fig:column_error_heatmap}b) exhibits the lowest overall errors. However, the highest error is observed at $\phi = 11.25\degree$ and $c = 70$ kPa, a cohesion value not included in the training data. For $a = 2.0$ (\Cref{fig:column_error_heatmap}c), higher errors appear in the low friction and cohesion cases, as already discussed in \Cref{fig:column_runout_comparison}. For $a = 0.5$ (\Cref{fig:column_error_heatmap}a), moderately elevated errors appear in some low cohesion cases (c = 10-25 kPa), but no distinctive trends exist. The model generalizes well to untrained material properties, such as $\phi = 45\degree$ and $c = 70$ kPa, except for a few outliers (e.g., $a = 1.0$, $\phi = 11.25\degree$, $c = 70$ kPa). Although the fine-tuning dataset excludes friction angles above $\phi=35\degree$, the model likely benefited from prior exposure to high-friction scenarios during base training. This result suggests that the fine-tuned GNS retains previously learned representations. Overall, the model maintains reliable accuracy within the trained ranges of $\phi$ and $c$, while demonstrating reasonable extrapolation performance in untrained regions. Most test cases result in smaller errors than the visual example shown in \Cref{fig:column_collapses}c, which still captures the overall runout profile.

% Need to answer
% - How does the error behaves with time for active particles?
% - Error is because of (1) distribution shift? (2) just highly mobile particles

\begin{figure}[!htbp]
    \centering
    \includegraphics[width=1.0\textwidth]{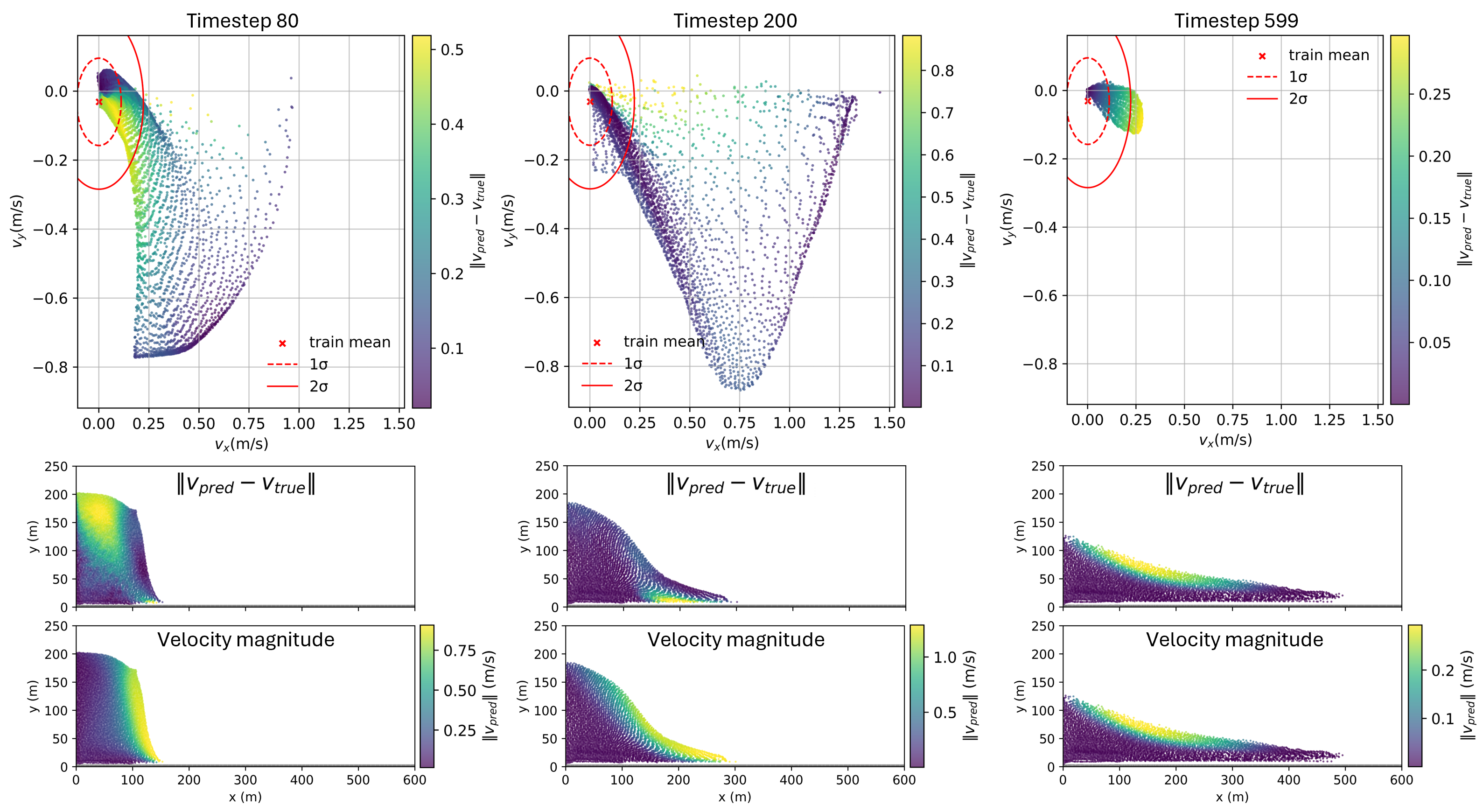}
    \caption{Velocity- and physical-space diagnostics of rollout errors at $t=80$, $200$, and $599$ for the column collapse test with a = 2.0, $\phi=11.25 \degree$, $c=10$ kPa. \textbf{Top row:} Predicted particle velocities $(v_x, v_y)$ colored by the velocity error magnitude $\lVert \mathbf{v}_{\mathrm{pred}}-\mathbf{v}_{\mathrm{true}}\rVert$; the red marker and contours show the mean, $1\sigma$, and $2\sigma$ ellipses of the fine-tuning data distribution. \textbf{Middle row:} GNS particle positions $(x,y)$ colored by $\lVert \mathbf{v}_{\mathrm{pred}}-\mathbf{v}_{\mathrm{true}}\rVert$. \textbf{Bottom row:} GNS particle positions colored by the predicted velocity magnitude $\lVert \mathbf{v}_{\mathrm{pred}}\rVert$.}
    \label{fig:error_distribution_ood}
\end{figure}

\Cref{fig:error_distribution_ood} presents velocity-space and physical-space diagnostics of rollout errors for the tall column collapse case associated with the largest runout error ($a = 2.0$, $\phi = 11.25^\circ$, $c = 10$ kPa). At the onset of flow mobilization ($t = 80$), the most active particles are concentrated near the developing flow front, as indicated by the velocity magnitude (bottom row). Despite their high mobility, velocity errors in this region remain relatively small (middle row). Instead, the largest errors are associated with particles undergoing near-vertical, free-fall-like motion. Importantly, particles that lie outside the core training velocity distribution (top row; red contours representing standard deviation with mean value) do not systematically exhibit larger errors; elevated errors are confined to a limited subset of particles experiencing near-free-fall dynamics. At the intermediate stage ($t = 200$), highly mobile particles remain concentrated at the flow front, while elevated errors localize primarily near the basal region, where rapid changes in velocity occur. By the final stage ($t = 599$), as the system approaches equilibrium, particle mobility and spatial error patterns align, and particles with velocities farther from the training mean, where in this case there is more mobility, exhibit higher errors.%decreases and velocity errors are substantially reduced across the domain, with only moderately higher errors for particles farther from the training mean.
Overall, these diagnostics indicate that increased errors arise primarily from flow regimes that are weakly represented in the training data, such as the near-free-fall regime observed at early times, and from particle states whose velocities move outside the training distribution at later stages. High particle mobility alone does not systematically lead to larger errors; rather, error growth is associated with mobility combined with out-of-distribution dynamics.

\subsubsection{Multi-layered slope runout}
\begin{figure}[!htbp]
    \centering
    \includegraphics[width=1.0\textwidth]{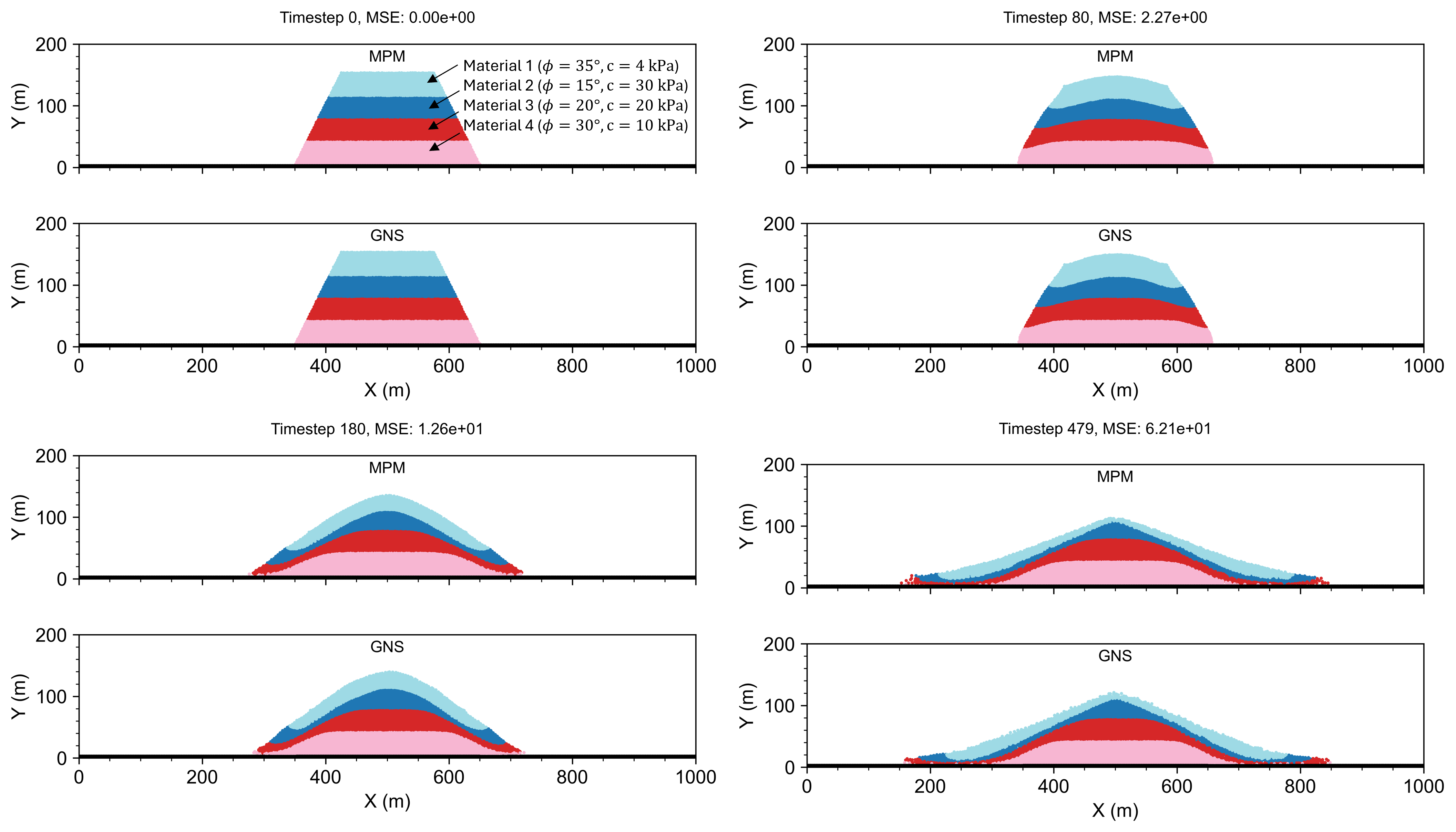}
    \caption{Flow evolution in a layered slope system simulated using the MPM and GNS. The slope consists of four materials with varying properties, labeled with different colors. The evolution is illustrated at four timesteps: 0, 80, 180, and 479. The material property values are shown in the figure at the timestep 0.}
    \label{fig:embankment_deposition}
\end{figure}

We evaluate the GNS model performance on the multi-layered slope system presented in \Cref{fig:embankment_deposition} (timestep 0 shows the initial configuration).
The system consists of four layers with different material properties. The width of the slope is 300 m, and the height is 150 m. The simulation domain spans 1000 m horizontally and 200 m vertically, exceeding the training domain dimensions (300 × 152 m or 400 m × 152 m). Additionally, the slope's four interacting layers also surpass the three-layer limit of the training configurations (refer to \Cref{table:training_datasets} and \Cref{fig:train_data_config}).

The first layer (material 1, top layer) has a friction angle $\phi = 35 \degree$ and cohesion $c = 4 \ \text{kPa}$. The second layer (material 2, the layer below the first layer) has a friction angle $\phi = 15 \degree$ and cohesion $c = 30 \ \text{kPa}$. The third layer (material 3, the layer below the second layer) has a friction angle $\phi = 20 \degree$ and cohesion $c = 20 \ \text{kPa}$. The fourth layer (material 4, the bottom layer) has a friction angle $\phi = 30 \degree$ and cohesion $c = 10 \ \text{kPa}$. The subsequent subfigures (\Cref{fig:embankment_deposition}) show the failure evolution with time at key timesteps 0, 80, 180, and 479. 

We first examine the evolution of the material distribution and runout patterns \Cref{fig:embankment_deposition}. At timestep=80, the system shows the onset of large movements. The edges of the layers start to displace outward while maintaining their layered structure. The inner parts experience minimal movement. These deformation patterns appear nearly identical between the MPM and GNS.

By timestep=180, the flow progresses further, with significant spreading and deformation observed in both simulators. The layered structure is preserved in both MPM and GNS, and the materials continue to displace outward, with the edges of each layer leading the deformation. Both simulations predict similar geometries, with the materials maintaining their relative positions and layering.

At the final timestep, the flow reaches equilibrium. Both simulators exhibit a similar flow profile with the layered structure preserved. The materials show a stretched and flattened bell-shaped geometry, with the third layer (Material 3, red-colored) undergoing the greatest lateral displacement, while the bottom material (Material 4, pink-colored) experiences less movement, entrapped by the other upper layers. Only the flanks of the bottom layer deform, with the inner central part experiencing less displacement. This deformation pattern is captured by the GNS simulations. The intermediate layers (materials 2 and 3) form the thin tail layers between the top and bottom layers along both ends of the deposit. The GNS successfully captures this subtle layer formation with a thickness similar to that of MPM. The results presented in this section highlight the GNS's ability to simulate multi-layered material interactions. The runout error estimated based on x = 500 m (symmetric x location) is 1.57\% on the left side of the flow and 0.96\% on the right side of the flow, again highlighting the adequate GNS's performance.

\begin{figure}[!htbp]
    \centering
    \includegraphics[width=1.0\textwidth]{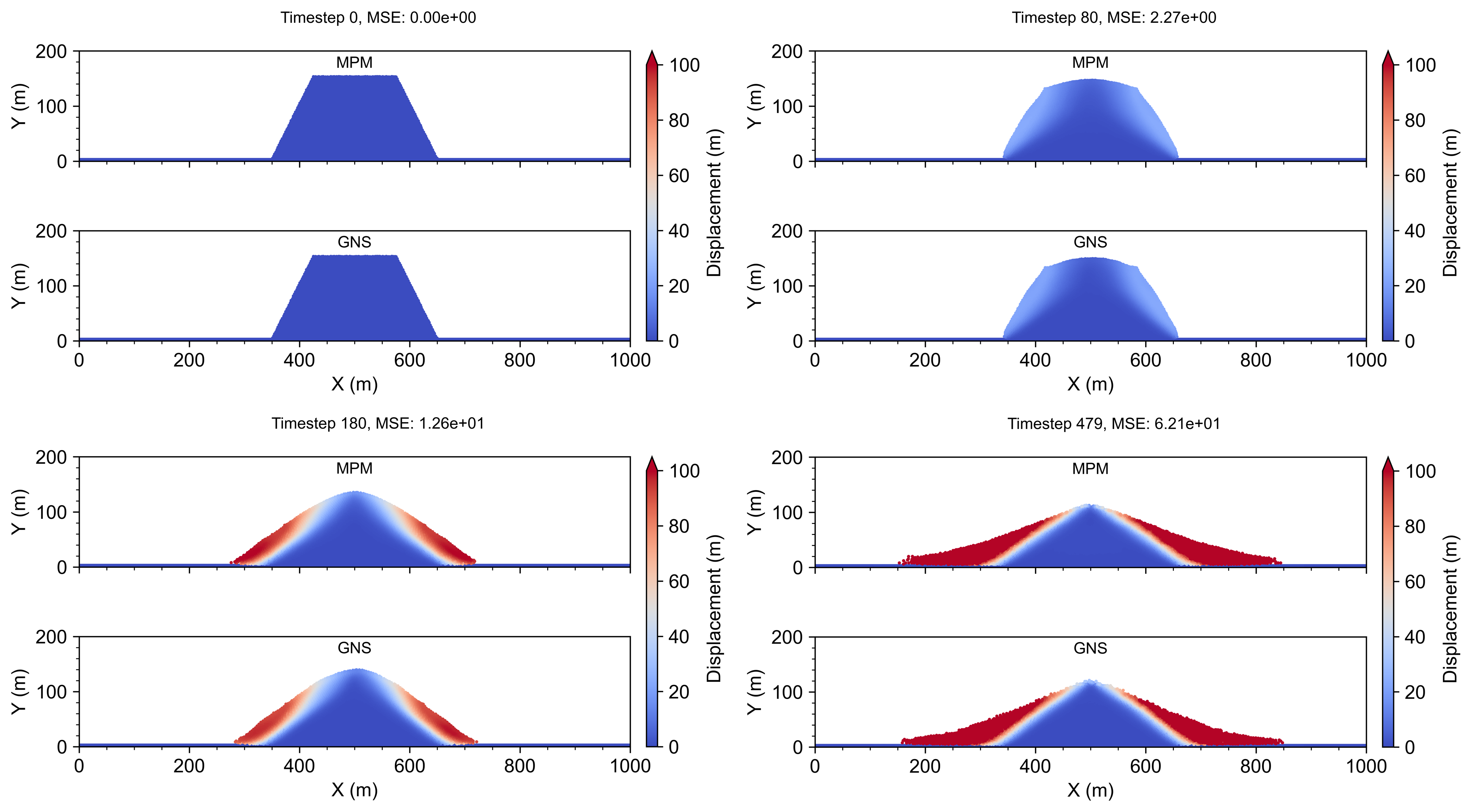}
    \caption{Evolution of granular flow of a dam-shaped multi-layered slope system simulated from the MPM and GNS. The color field represents displacement magnitude. The evolution is illustrated at four timesteps: 0, 80, 180, and 479.}
    \label{fig:embankment_disp}
\end{figure}

\Cref{fig:embankment_disp} shows the displacement field evolution associated with \Cref{fig:embankment_deposition}. At timestep=80, the GNS captures the collapse development well. The highest displacement is concentrated along the outermost edges, shown in light blue, while the central region remains stationary in dark blue. The GNS successfully tracks the intermediate progress of the flow (see timestep=180), with the highest displacement continuing to occur at the slope's flanks. At timestep 479, the last timestep when the flow reaches equilibrium, the displacement field of the GNS closely aligns with MPM, with a clear concentration of displacement on the left and right slopes. Finally, we also assess the computational efficiency of the GNS relative to MPM for the configuration shown in \Cref{fig:embankment_deposition}. MPM simulations were performed on the Frontera supercomputer at the TACC using an Intel Cascade Lake processor with 56 cores, while the GNS simulations were executed on an NVIDIA Quadro RTX 5000 GPU with 16 GB of memory. The GNS completed the simulation in just 41 seconds, compared to 6,075 seconds for MPM, achieving a speed-up of approximately 145×.

% Talk about the same-device comparison

To address the hardware dependence of this comparison, we additionally evaluated GNS inference on the same CPU hardware used for the MPM simulations. Under this configuration, the GNS required 2,001 seconds, yielding a speed-up of approximately 3× relative to the CPU-based MPM simulation. While this same-hardware comparison highlights the intrinsic efficiency of the learned simulator, the 145× speed-up more accurately reflects practical deployment scenarios, in which GNS models naturally benefit from GPU acceleration. For context, prior studies \citep{choi2024graph} report that GPU-accelerated MPM implementations \citep{arduino2021tsunami} achieve speed-ups of approximately 5-10× relative to the CPU-based CB-Geo MPM implementation used here. Even under such improved MPM performance, the GNS framework retains a meaningful computational advantage. \cite{zhao2025physical} provides further evidence of this by comparing MPM and GNS on the same GPU device and reporting a speed-up of about 100x.

% Error accumulation

\begin{figure}[!htbp]
    \centering
    \includegraphics[width=0.5\textwidth]{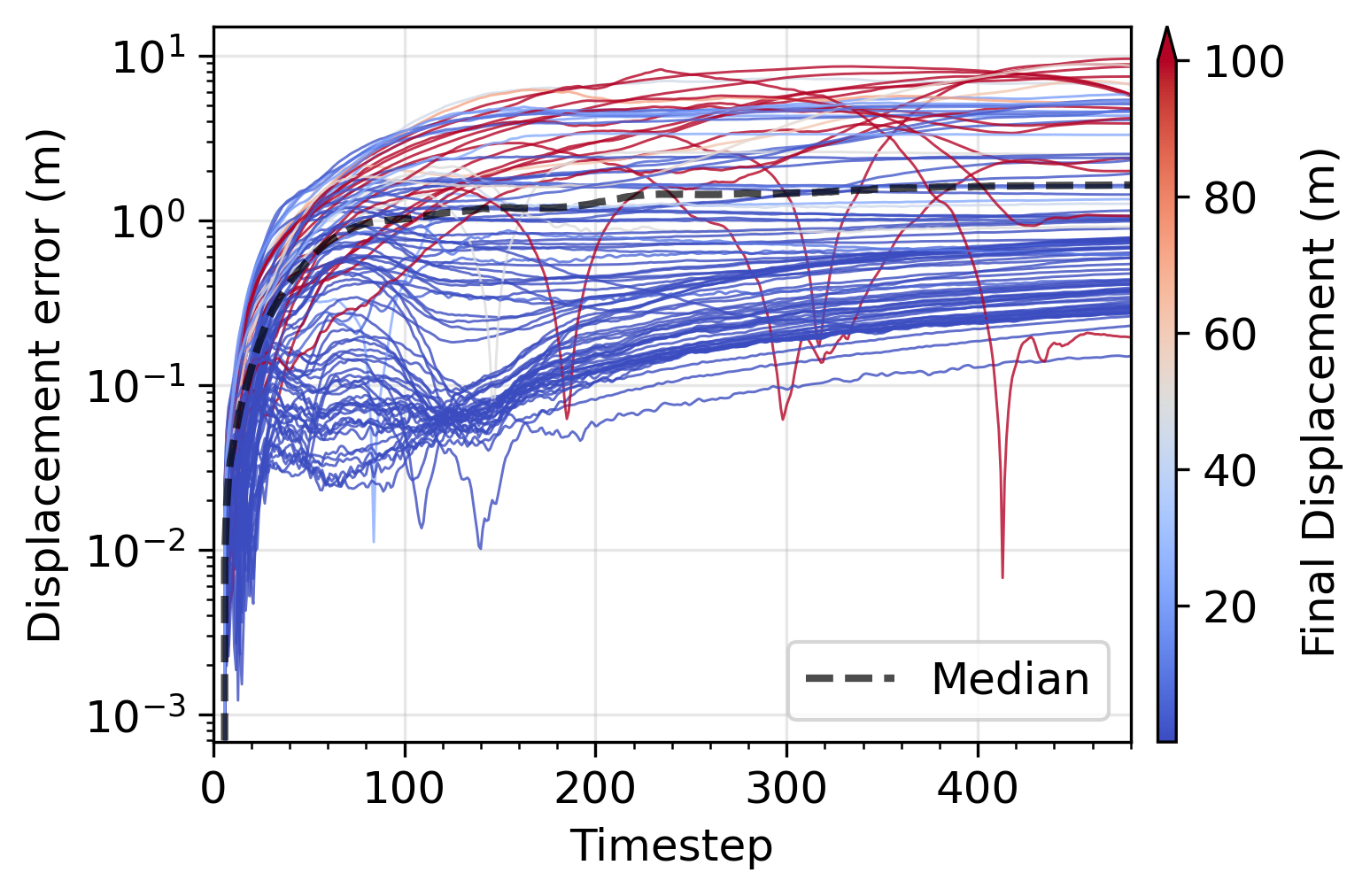}
    \caption{Median displacement error evolution over time (black dashed line) for the material points shown in \Cref{fig:embankment_disp}. Colored lines represent individual particle error histories, with line colors indicating final runout displacements.}
    \label{fig:embankment_error_accumulation}
\end{figure}

\Cref{fig:embankment_error_accumulation} shows the evolution of median displacement error over time (black dashed line) for the material points analyzed in \Cref{fig:embankment_disp}. To illustrate particle-level variability in error accumulation, we also plot the displacement error histories of 100 randomly selected particles, with line color indicating each particle’s final runout distance. As expected from the autoregressive nature of GNS predictions, displacement errors accumulate over time, particularly for particles undergoing large displacements. These particles, primarily located along the flanks of the slope (\Cref{fig:embankment_disp}), follow longer, more dynamic trajectories and are more susceptible to cumulative error. Despite this, the GNS model still captures the overall runout pattern observed in the MPM model. In contrast, particles with limited motion, typically located within the slope body, maintain consistently low error throughout the simulation. As the slope approaches equilibrium and particle motion ceases, error growth stabilizes, and displacement errors plateau. Similar trends have been reported in prior GNS studies of small-scale granular column collapse using frictional materials \citep{zhao2025physical,choi2024graph}.

\subsection{Inverse analysis}

\begin{figure}[!htbp]
    \centering
    \includegraphics[width=1.0\textwidth]{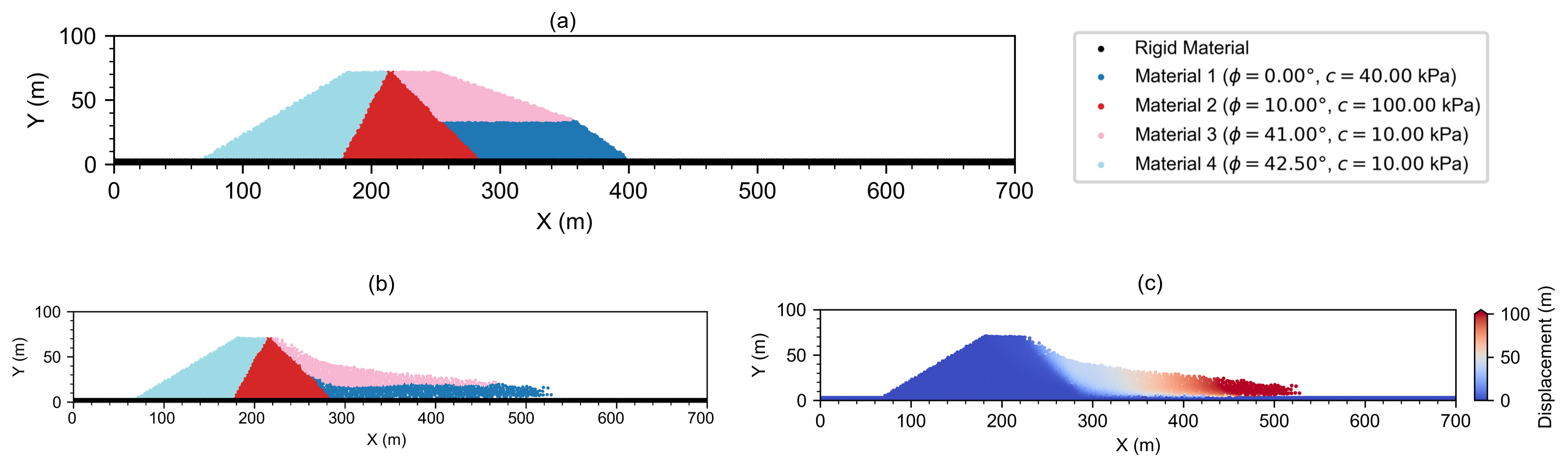}
    \caption{Multi-material dam considered in the inverse analysis. (a) initial configuration, (b) final deposit after failure, (c) displacement field of the final deposit.}
    \label{fig:inverse-initial-geometry}
\end{figure}

This section demonstrates the application of the differentiable GNS framework (see \Cref{fig:differentiable_gns}) to perform inverse analyses for a dam runout scenario. We use the trained GNS model developed in this study to back-calculate the material strength parameters for a multi-material dam system based on the final runout geometry. The initial configuration for the inverse analysis appears in \Cref{fig:inverse-initial-geometry}a and is not part of the training dataset.

The dam consists of three vertical zones: (1) the upstream shell (sky blue), (2) the core (red), and (3) the downstream shell, which includes two horizontal layers (pink and blue). \Cref{table:inverse_material_config} summarizes the true material properties used to generate the runout via MPM. Material 1, representing part of the downstream shell, has an undrained strength of $s_u=40 \ \text{kPa}$, modeled using Mohr-Coulomb parameters $\phi=0 \degree$, $c=40 \ \text{kPa}$. This strength corresponds to an undrained strength ratio $s_u / \sigma_v'=0.13$, based on an average depth of 17 m. Material 2 (core) has $\phi=10 \degree$, $c=100 \ \text{kPa}$; material 3 has $\phi=41 \degree$, $c=10 \ \text{kPa}$; and material 4 (upstream shell) has $\phi=42.5 \degree$, $c=10 \ \text{kPa}$. Notably, several of these values fall outside the GNS training range for friction angle ($\phi=3 \degree$ to $\phi=40 \degree$) allowing us to evaluate generalization performance.

\resizebox{\textwidth}{!}{%
\begin{threeparttable}
\centering
\caption{Material properties and optimization settings for the multi-material dam in \Cref{fig:inverse-initial-geometry}. The table lists the true values of friction angle ($\phi$) and cohesion ($c$), indicates whether each parameter is optimized or fixed in the inverse analysis, and reports the lower and upper bounds used by the L-BFGS-B optimizer.}
\label{table:inverse_material_config}
\begin{tabular}{@{}ccccccccc@{}}
\toprule
\multirow{2}{*}{Materials} & \multirow{2}{*}{$\phi$} & \multirow{2}{*}{$c$} &
\multicolumn{3}{c}{$\phi$ ($\degree$)} &
\multicolumn{3}{c}{$c$ (kPa)} \\ 
\cmidrule(l){4-9}
 & & & True & Min constraint & Max constraint & True & Min constraint & Max constraint \\ 
\midrule
 1 & Fixed & Optimized & 0 & N/A & N/A & 40\tnote{a} & 10 & 100 \\
 2 & Fixed & Optimized & 10 & N/A & N/A & 100 & 99 & 101 \\
 3 & Optimized & Fixed & 41 & 5 & 50 & 10 & N/A & N/A \\
 4 & Fixed & Fixed & 42.5 & N/A & N/A & 10 & N/A & N/A \\ 
\bottomrule
\end{tabular}

\begin{tablenotes}\footnotesize
\item[a] For material 1, we assign $\phi = 0 \degree$ and $c = 40 \ \text{kPa}$ to represent its ``true'' undrained shear strength $s_u = 40 \ \text{kPa}$. This is derived from the assumed shear strength ratio $s_u / \sigma_v'$ of 0.13 at the average depth of material 1 (17 m).
\end{tablenotes}
\end{threeparttable}}

We use MPM to generate the target final deposit, shown in \Cref{fig:inverse-initial-geometry}b-c, which serves as the reference for optimization. The objective is to optimize $c$ (i.e., undrained shear strength) of material 1 and $\phi$ of material 3 so that the post failure geometry from the GNS prediction matches the MPM-derived post failure geometry. In addition, we impose tight bounds on the cohesion $c$ of material~2 to evaluate whether L-BFGS-B properly enforces strict constraint ranges. The remaining material properties are treated as known and are therefore not included in the optimization. \Cref{table:inverse_material_config} summarizes the optimized strength parameters and their associated bounds. Although conventional back-analyses in engineering practice and post-failure forensic investigations \citep{olson2001liquefaction} often focus on a single parameter, this study simultaneously back-calculates three parameters, which poses a more challenging inverse problem.

We define the inverse loss as the mean squared error (MSE) between the GNS-predicted particle positions $\hat{P_{\theta}^t}$ and the target positions $P_{\theta}^t$ from the MPM simulation, using the formulation in \Cref{eq:inverse_loss}. Since we only consider the final timestep, we set $|\mathcal{T}|=1$.

\begin{figure}[!htbp]
    \centering
    \includegraphics[width=0.9\textwidth]{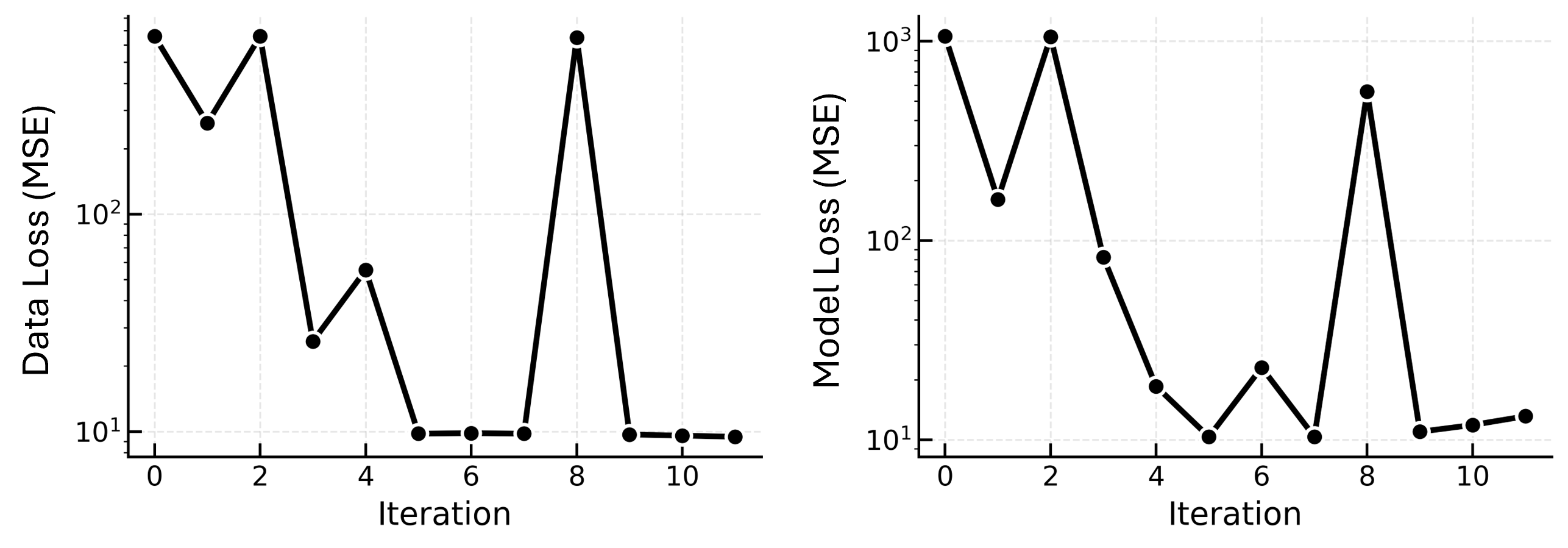}
    \caption{Loss history during optimization. At each iteration, data loss is computed with \Cref{eq:inverse_loss} where the ground truth particle positions $P_{\theta}^t$ is the final deposit obtained from the MPM simulation, and $\hat{P_{\theta}^t}$ is the GNS simulation result with the current material model parameters. The model loss represents the MSE between the normalized true material model properties and the inferred material properties at the current iteration.}
    \label{fig:opt_history}
\end{figure}

\Cref{fig:opt_history} shows the optimization history. The left panel tracks data loss (MSE between particle positions), and the right panel tracks model loss (MSE between normalized true and inferred parameters). Both metrics converge after around 10 iterations with decreasing trends.

\begin{figure}[!htbp]
    \centering
    \includegraphics[width=1.0\textwidth]{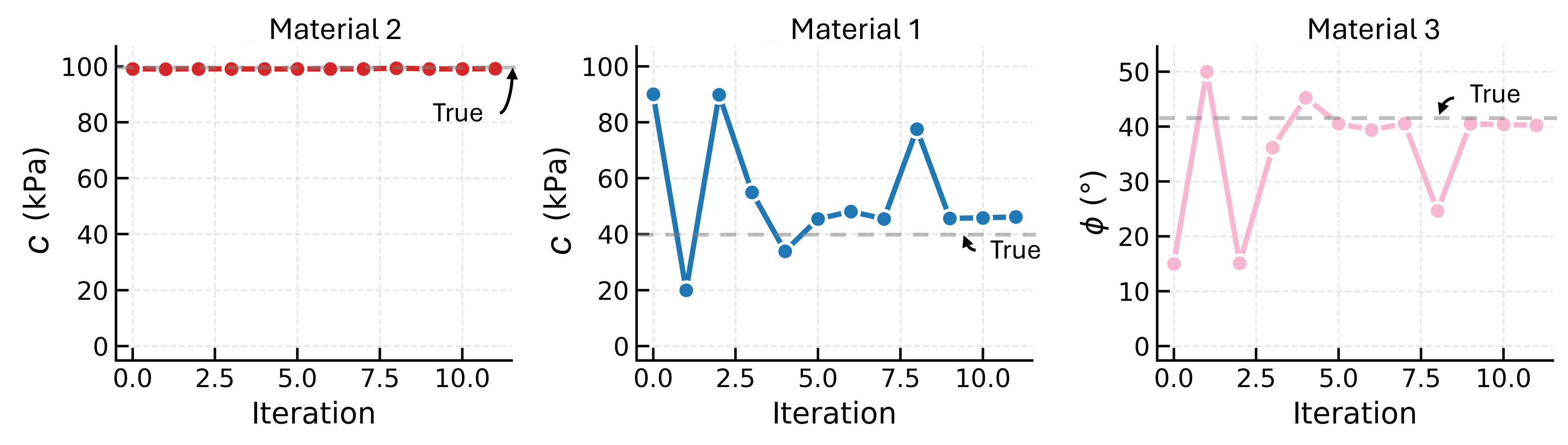}
    \caption{Material model parameter history during the optimization.}
    \label{fig:opt_history_materials}
\end{figure}

\Cref{fig:opt_history_materials} shows the change in material parameters during inverse optimization. The cohesion of material 1 (blue dashed line) converges to $46.17 \ \text{kPa}$, which is close to the true value of $40 \ \text{kPa}$. The estimated strength corresponds to a strength ratio $s_u / \sigma_v'=0.15$, which is close to the target value of 0.13. This small discrepancy ($\sim 0.02$) falls within the uncertainty typically encountered in empirical back-analyses \citep{olson2003yield}. Throughout the optimization, material 2's parameters remain within the imposed bounds, demonstrating that the L-BFGS-B algorithm effectively maintains constraints while achieving convergence. Material 3's $\phi$ converges to $40.24\degree$, showing good agreement to the true value of $41\degree$.

\begin{figure}[!htbp]
    \centering
    \includegraphics[width=1.0\textwidth]{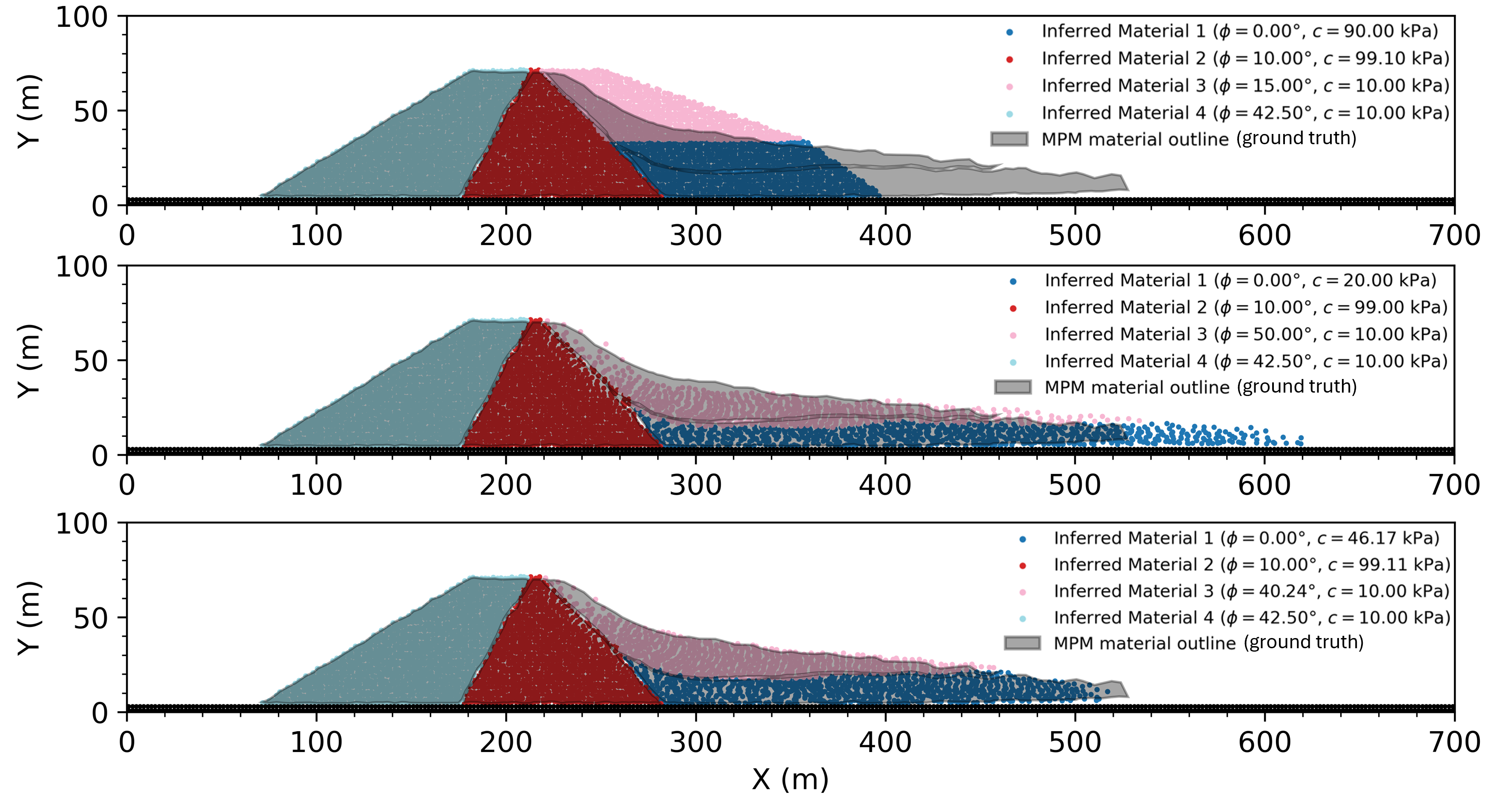}
    \caption{Visualization of the GNS prediction during optimization compared to the final deposit from MPM. The legend shows the inferred material properties at each iteration. The black shaded outlines correspond to the final deposit from MPM for each material.}
    \label{fig:inverse-geometric-history}
\end{figure}

\Cref{fig:inverse-geometric-history} illustrates how the estimated runout geometries change during the optimization process. At iteration 0, the GNS predicts no deformation. At iteration 1, the optimizer reduces material 1's $c$ to around $20 \ \text{kPa}$, causing excessive runout. As the optimization proceeds and converges, the GNS-predicted geometry progressively approaches and ultimately matches the target runout obtained from the MPM simulation. 

It is worth noting that the inversion loss function is flexible in its definition. In the example discussed here, the loss function is defined based on final runout geometry (\Cref{eq:inverse_loss}); however, alternative formulations are possible. For instance, higher weights could be assigned to particles experiencing larger displacements, such as those near the runout front. For the case considered, the current loss function performs well, as the inverted model reproduces the target runout geometry without requiring additional weighting. Another consideration is that inversion based solely on the final geometry may not yield a unique solution. However, in practical applications, only the final runout geometry is typically available \citep{kramer2015empirical,yerro2019runout_mpm_oso,olson2001liquefaction}. As a result, domain expertise plays an important role in assessing the plausibility of the inverted parameters. When additional information is available, such as intermediate deformation geometries (i.e., $|\mathcal{T}|>1$ in \Cref{eq:inverse_loss}), it can be incorporated to further constrain the inversion.

Regarding computational cost, each GNS-based optimization step takes $\sim 1$ minute on an NVIDIA Quadro RTX 5000, including 20 seconds for forward simulation and 42 seconds for gradient computation. In contrast, a single iteration using a non-differentiable forward model (e.g., MPM) would require over 95 minutes, including 2850 seconds for one MPM run, plus three additional forward simulations for gradient estimation via finite differences for each of three optimizing material parameters, which are also known for introducing numerical noise. In this context, the computational efficiency of the proposed inversion is remarkable.

\section{Discussion: Data size sensitivity and future data generation strategies}

\subsection{Data size sensitivity}

\begin{figure}[!htbp]
    \centering
    % Subfigure (a)
    \begin{subfigure}[b]{0.48\textwidth}
        \centering
        \includegraphics[width=\textwidth]{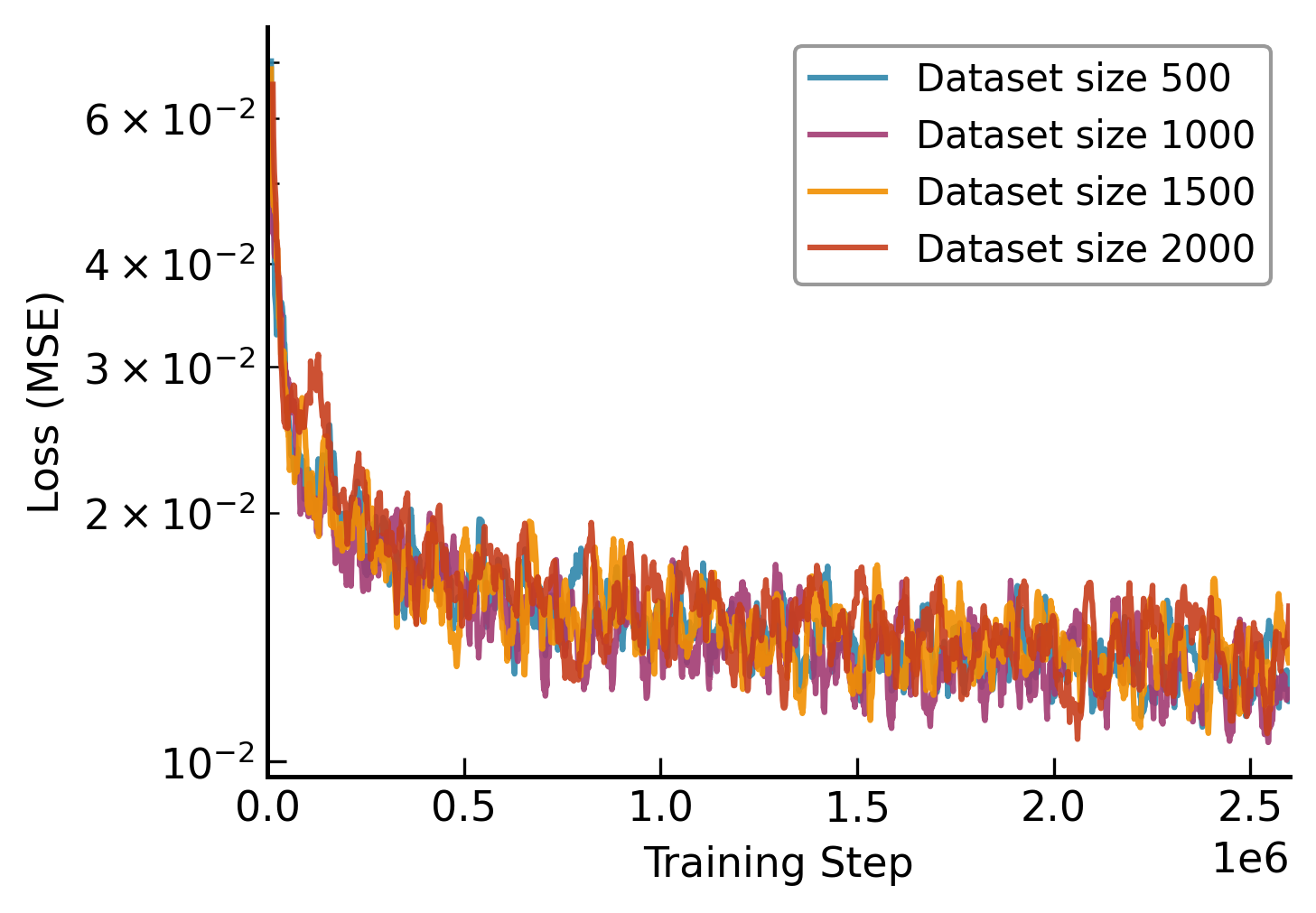} 
        \caption{}
        \label{fig:loss_history_base}
    \end{subfigure}
    \hfill
    % Subfigure (b)
    \begin{subfigure}[b]{0.48\textwidth}
        \centering
        \includegraphics[width=\textwidth]{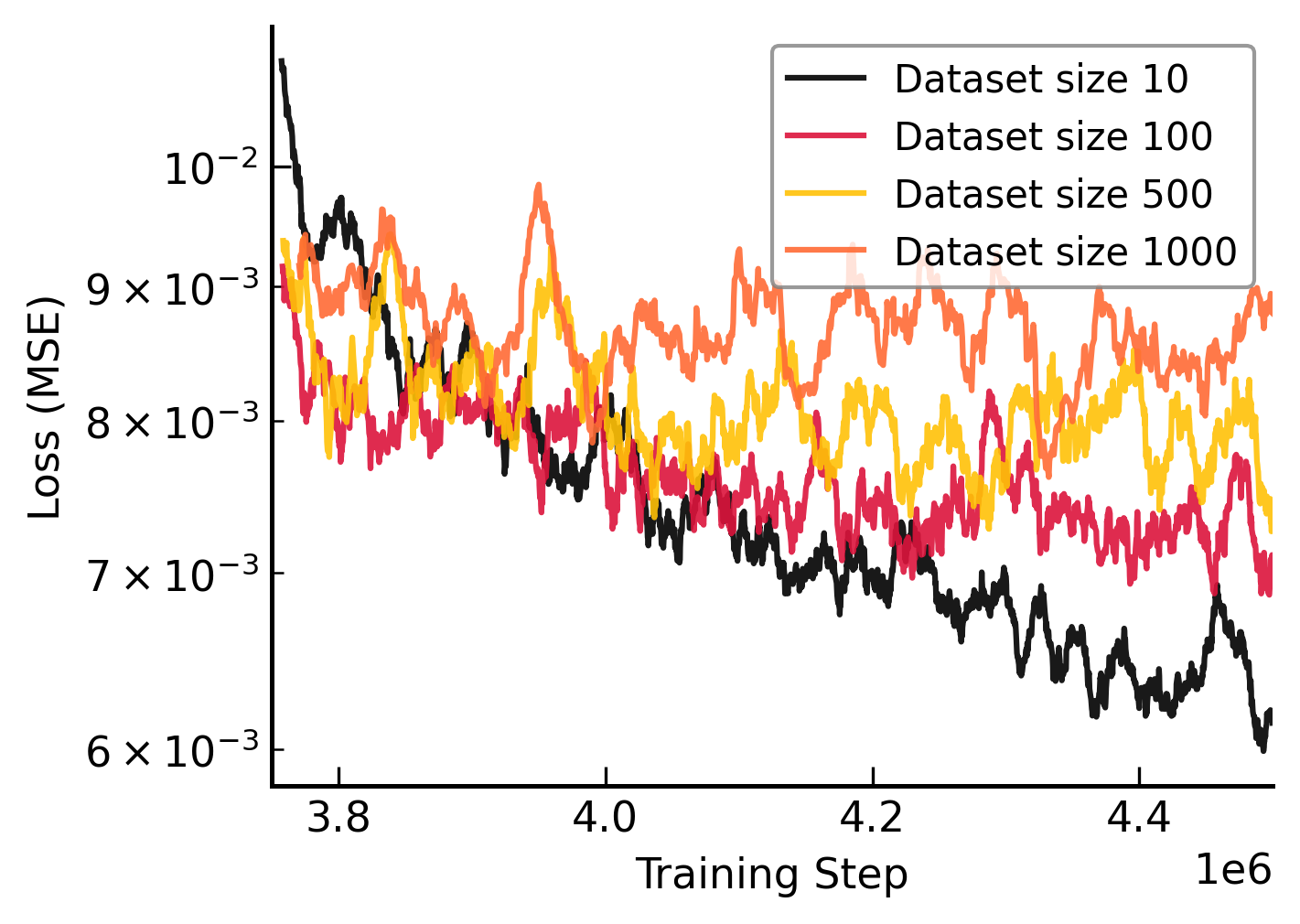} 
        \caption{}
        \label{fig:loss_history_finetuning}
    \end{subfigure}
    
    \caption{The loss history with different dataset sizes for (a) base model training and (b) fine-tuning model training.}
    \label{fig:data_sensitivity_training_loss}
\end{figure}

We evaluate the sensitivity of GNS performance to dataset size during both the base training and fine-tuning phases. For base model training, we consider four dataset sizes: N = 500, 1000, 1500, and 
2000 simulations (the full base dataset). For fine-tuning, we consider N = 10, 50, 100, 250, 500, 750, and 
1000 simulations (the full fine-tuning dataset). \Cref{fig:data_sensitivity_training_loss}a shows the training loss histories for the base models trained for 2.6 million steps using identical hyperparameters. All dataset sizes exhibit similar loss-decay trends, indicating that the model can learn the dominant runout dynamics with datasets as small as 500 simulations under a fixed training-step budget.

\Cref{fig:data_sensitivity_training_loss}b shows the fine-tuning loss histories, where all models were fine-tuned up to 4.5 million steps following the base model training with N = 2000. Models fine-tuned with smaller datasets exhibit faster loss reduction under a fixed-step budget, which is attributed to their experiencing more effective epochs (i.e., more passes through the data). This behavior reflects a compute-limited training regime rather than improved data efficiency.

\begin{figure}[!htbp]
    \centering
    \includegraphics[width=1.0\textwidth]{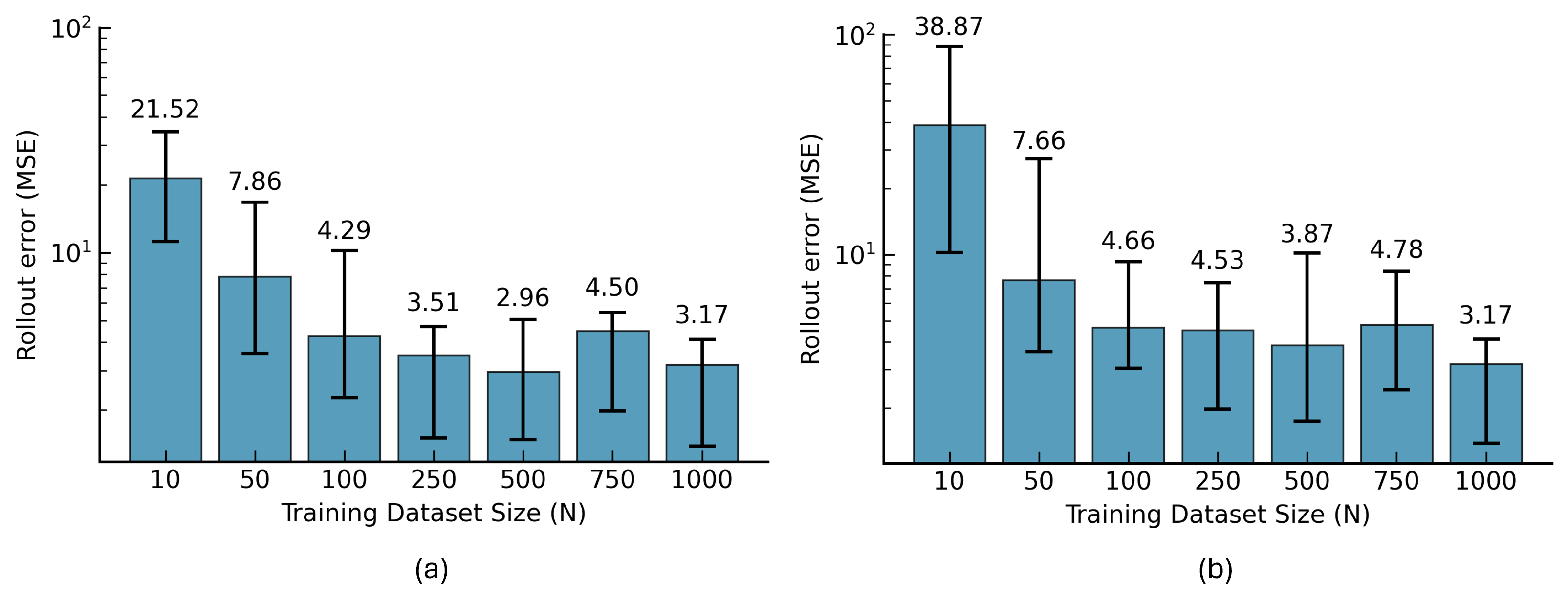}
    \caption{The rollout error (MSE) for the fine-tuning model with different dataset sizes evaluated at (a) the 4.5 million steps and (b) 10 epochs. The bars show the median values with error bars indicating the 25th-75th percentile ranges.}
    \label{fig:rollout_mse_finetune}
\end{figure}

To assess generalization performance, we evaluate the fine-tuned models on rollout tasks using a test set of 20 unseen simulations derived from the configurations shown in \Cref{fig:train_data_config}. \Cref{fig:rollout_mse_finetune}a reports rollout errors under a fixed budget of 4.5 million training steps, isolating the effect of dataset size under a constant number of parameter updates. \Cref{fig:rollout_mse_finetune}b reports results for a fixed-epoch setting (10 epochs), corresponding to approximately 4.5 million steps for 
N = 1000, in which the number of passes through the data is held constant while total compute increases with dataset size. The two scenarios show a rollout error increase when the dataset size drops below approximately 50-100 simulations, indicating limited generalization when the training distribution is undersampled. For dataset sizes of 250 and above, the median rollout error stabilizes. The fixed-epoch setting (\Cref{fig:rollout_mse_finetune}b), generally exhibits larger errors than in the fixed-step case, likely because of the reduced number of parameter updates.

Overall, these results indicate that, for the tasks considered here, moderate dataset sizes on the order of 100-250 simulations can achieve a performance comparable to that obtained with larger datasets, provided that sufficient training compute is available. This finding suggests that meaningful reductions in dataset-generation cost may be possible without substantial loss in accuracy for similar problem settings. However, the results cannot be generalized; a comprehensive investigation of the trade-offs among dataset size, training budget, and generalization performance across broader task distributions is beyond the scope of this study and represents an important direction for future work.

\subsection{Data generation}
Although this study focused on the configurations shown in \Cref{fig:train_data_config}, which target representative runout responses in layered slope systems, future work could adopt more sophisticated data generation strategies. These include incorporating geometries derived from case histories of slope failures as well as from well-performing slope systems. Prioritization will be necessary, as generating a large number of geometric variations would be computationally prohibitive. One potential direction is the use of active learning frameworks (e.g., \cite{yuan2024active_learning_tunel,siacara2024active_learning_slope}), where the model identifies high-uncertainty predictions to guide the selection of new training data. This could potentially enable more efficient and targeted data generation that focuses on the most informative cases. Another potential avenue involves generative models (e.g., \cite{parsa2025genai_geo,mosser2017genai_porous}) to automatically produce a broad range of plausible slope geometries. However, developing such models to generate realistic and physically meaningful configurations, and validating that these augmentations improve model performance, is expected to be a challenging task.

\section{Conclusion}
This study introduces a differentiable Graph Neural Network Simulator (GNS) for forward and inverse modeling of multi-layered slope systems characterized by the Mohr-Coulomb model. We demonstrate that the GNS serves as an effective differentiable surrogate model for the Material Point Method (MPM), significantly accelerating both forward and inverse modeling of granular flow simulations in realistic multi-layered slope systems while considering granular flow evolution.

Through transfer learning, we fine-tuned a pre-trained GNS model, originally trained only for a single material property (friction angle), to incorporate cohesion, rather than training a completely new model from scratch. The fine-tuning training loss successfully stabilizes when facing previously unseen granular flow data involving cohesive materials.

For forward modeling, we first evaluate GNS's performance to simulate Mohr-Coulomb-type materials using granular column collapse experiments. The results show that the GNS adequately captures the runout behavior across diverse friction angles and cohesion values. There are larger errors when simulating highly dynamic fluid-like materials outside the training dataset range, but these cases are not common in engineering applications. In general, the errors remain below 10 percent for most scenarios. We then apply the GNS to simulate a multi-layered dam runout. The GNS accurately reproduces runout dynamics and inter-material interactions, while achieving up to 145× speed-up compared to MPM.

For inverse modeling, we apply the differentiable GNS framework to identify material parameters in a multi-layered slope. By combining the GNS-accelerated forward simulation, L-BFGS-B optimization, and reverse-mode automatic differentiation, the proposed approach successfully identified the undrained shear strength of a selected material in a multi-material dam system within minutes. An equivalent inverse analysis using conventional MPM for forward evaluation and finite differentiation for gradient computation would require several hours of computational time. Although we focus on a single-parameter inversion, because it is the common scenario in back-analyses of slope failures, the proposed framework supports efficient multi-parameter inversion. This is the case because the reverse-mode AD computes exact gradients for all parameters in a single forward pass, keeping the computational cost nearly constant regardless of parameter count. In contrast, finite difference methods, which would be required with MPM schemes, scale poorly, as they require multiple forward evaluations per parameter.

We also found that, for the fine-tuning, performance is comparable across moderate and large training datasets when evaluated under both compute-limited (fixed training steps) and fixed-exposure (fixed epoch) regimes, provided that the dataset size remains above approximately $N \approx 50 \ \text{to} \ 100$ simulations. Below this range, performance degrades, indicating insufficient coverage of the training distribution. For dataset sizes with N higher than about 250, the rollout error stabilizes and remains close to that obtained with substantially larger datasets. These results suggest that, for the tasks considered in this study, moderate fine-tuning datasets can achieve performance comparable to larger ones, up to a threshold below which performance deteriorates. This finding highlights the potential to reduce the computational cost of data generation for fine-tuning, while also underscoring the importance of maintaining a sufficient dataset size. These findings are task-specific, and a comprehensive assessment of the trade-offs among dataset size, training budget, and generalization performance across broader problem classes lies beyond the scope of this study and represents a direction for future work.

Lastly, while we considered only the final runout geometry for inversion, the presented GNS framework can accommodate intermediate runout states, enabling more robust optimization. As field instrumentation and remote sensing continue to improve, future case histories are likely to capture the runout progression, making the proposed GNS framework especially well-suited for emerging applications.

This study demonstrates the potential of differentiable GNS as a useful surrogate modeling approach for forward and inverse problems in slope systems, effectively balancing computational efficiency with physical accuracy. 

This study focuses on 2D slope systems, which reflects the format of most available field data and standard geotechnical practice \citep{olson2001liquefaction,yerro2019runout_mpm_oso,kramer2015empirical,weber2015engineering}. However, extending the GNS to fully 3D geometries in realistic cases would be the future direction.

\section{Acknowledgment}

This study was funded by the TAILENG foundation. Any opinions, findings, conclusions, or recommendations expressed in this study are those of the author(s). We also thank the Texas Advanced Computing Center (TACC) at The University of Texas at Austin for providing Frontera and Lonestar6 HPC resources to support the GNS training (https://www.tacc.utexas.edu). This work was also supported by the InnoCORE program of the Ministry of Science and ICT (N10250154).




\bibliographystyle{elsarticle-harv} 
\bibliography{references}





% \end{thebibliography}
\clearpage
\appendix

\renewcommand{\thefigure}{A.\arabic{figure}} % Change figure numbering to A.1, A.2, ...
\setcounter{figure}{0} % Reset figure counter
\renewcommand{\thetable}{A.\arabic{table}}
\setcounter{table}{0}

\section{Training data}\label{sec:appendix_training_data}

\begin{table}[htbp]
\footnotesize
\centering
\caption{Coordinates of the polygon vertices in \Cref{fig:train_data_config}.}
\setlength{\tabcolsep}{10pt}
\renewcommand{\arraystretch}{0.95}
\begin{tabular}{@{}lll@{}}
\toprule
Polygon vertex indices & X (m) & Y (m) \\ \midrule

\multicolumn{3}{@{}l}{\textbf{\Cref{fig:train_data_config}a}} \\
{[0]} & 0 & 4 to 40 \\
{[1] to [3]} & 10 to 290 & 4 to 40 \\
{[4]} & 300 & 4 to 40 \\

\multicolumn{3}{@{}l}{\textbf{\Cref{fig:train_data_config}b}} \\
{[0]} & 0 & 4 \\
{[1]} & 125 to 193 & 4 \\
{[2]} & 137.5 to 225 & 4 \\
{[3]} & 150 to 225 & 4 \\
{[4]} & 0 & 12.5 to 50 \\
{[5]} & 100 & 12.5 to 50 \\
{[6]} & 0 & 25 to 75 \\
{[7]} & 100 & 25 to 37.5 \\
{[8]} & 0 & 50 to 100 \\
{[9]} & 100 & 50 to 100 \\

\multicolumn{3}{@{}l}{\textbf{\Cref{fig:train_data_config}c}} \\
{[0]} & 0 & 4 to 50 \\
{[1]} & 60 to 100 & 4 \\
{[2]} & 90 to 140 & 4 \\
{[3]} & 90 to 150 & 4 \\
{[4]} & 0 & 30 to 100 \\
{[5]} & 10 to 80 & 70 to 100 \\
{[6]} & 0 & 120 to 140 \\
{[7]} & 10 to 80 & 80 to 120 \\

\multicolumn{3}{@{}l}{\textbf{\Cref{fig:train_data_config}d}} \\
{[0]} & 0 to 110 & 0 \\
{[1]} & 0 to 120 & 0 \\
{[2]} & 180 to 300 (400 for fine-tuning dataset) & 0 \\
{[3]} & 210 to 300 (400 for fine-tuning dataset) & 0 \\
{[4]} & 0 & 4 to 40 \\
{[5]} & 10 to 390 & 4 to 40 \\
{[6]} & 10 to 390 & 4 to 40 \\
{[7]} & 400 & 4 to 40 \\
{[8]} & 40 to 170 & 30 to 60 \\
{[9]} & 130 to 260 & 30 to 60 \\
{[10]} & 30 to 160 & 50 to 190 \\
{[11]} & 140 to 270 & 50 to 90 \\

\bottomrule
\end{tabular}
\label{table:appendix_training_data}
\end{table}

\begin{figure}[!htbp]
    \centering
    \includegraphics[width=1.0\textwidth]{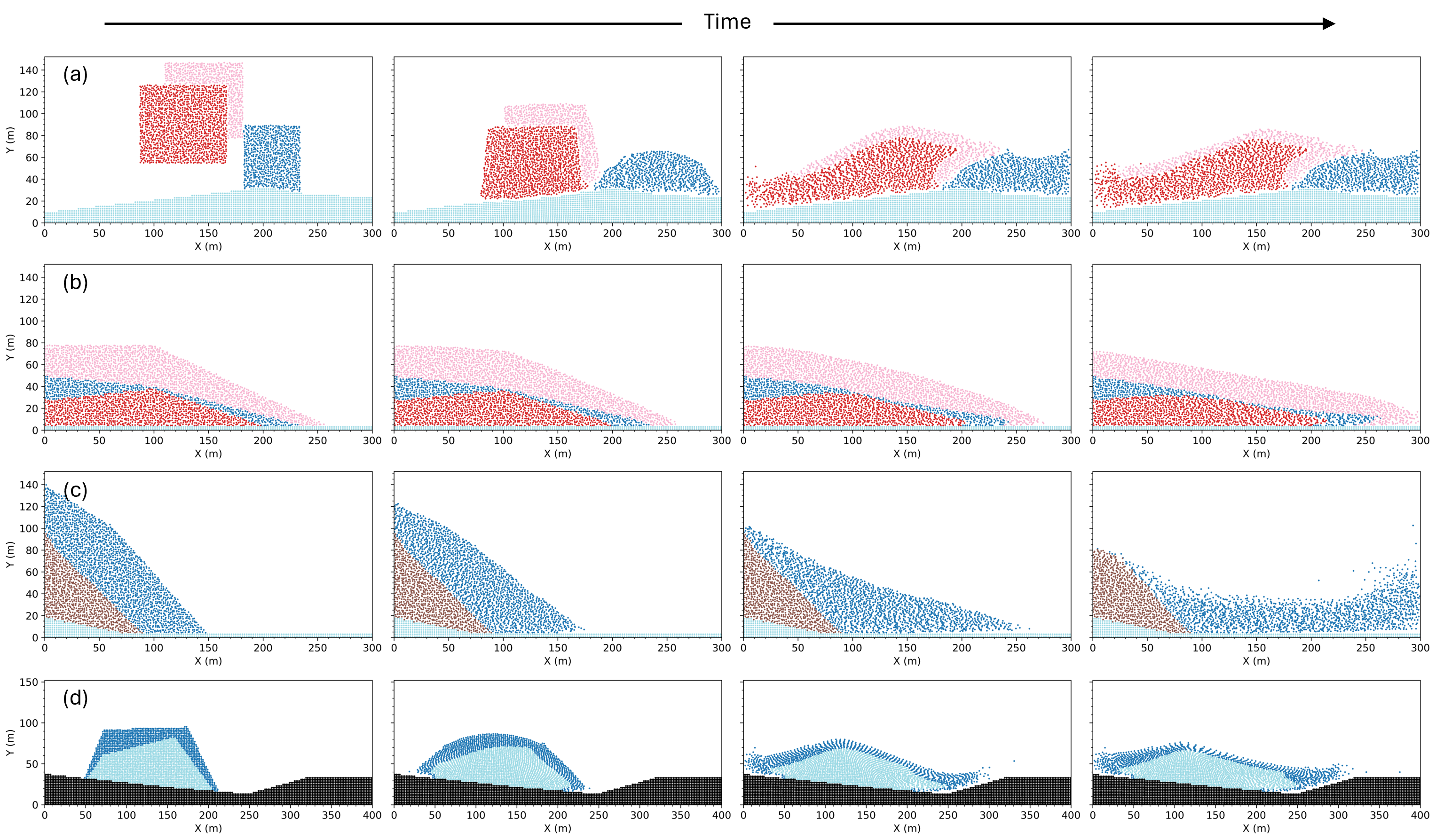}
    \caption{MPM simulation results used in training data (\Cref{fig:train_data_config}). Each subfigure is related to each subfigure in \Cref{fig:train_data_config}.}
    \label{fig:appendix_training_data}
\end{figure}

\end{document}